\documentclass[]{aastex}

\usepackage{graphicx,subfigure}

\def \lta {\mathrel{\vcenter
     {\hbox{$<$}\nointerlineskip\hbox{$\sim$}}}}
     
\newcommand{\Msun}{\ensuremath{M_{\odot}}}
\newcommand{\Mdot}{\ensuremath{\dot{M}}}
\newcommand{\MCh}{\ensuremath{M_{\rm Ch}}}
\newcommand{\MWD}{\ensuremath{M_{\rm WD}}}
\newcommand{\RWD}{\ensuremath{R_{\rm WD}}}

\newcommand{\viscKH}{\ensuremath{\nu_{\rm KH}}}
\newcommand{\viscBC}{\ensuremath{\nu_{\rm BC}}}
\newcommand{\viscZ}{\ensuremath{\nu_{\rm Z}}}
\newcommand{\verthat}{\ensuremath{{\bf \hat{r}}}}
\newcommand{\horihat}{\ensuremath{{\bf \hat{\theta}}}}
\newcommand{\zeehat}{\ensuremath{{\bf \hat{z}}}}
\newcommand{\azihat}{\ensuremath{{\bf \hat{\phi}}}}

\def \ie{{\sl i.e.}}
\def \eg{{\sl e.g.}}
\def \viz{{\sl viz.}}
\def \etal{{\sl et al.}}

\newcommand{\insertfig}[4]{
  \begin{figure}[!htbp]
   \begin{center}
      \includegraphics[#1]{#2}
   \end{center}
   \vspace{-0.00cm}
   \caption{#3}
   \label{#4}
  \end{figure}
}

\newcommand{\insertdoblfig}[6]{
  \begin{figure}[!htbp]
   \begin{center}
     \includegraphics[#1]{#2}
     \includegraphics[#3]{#4}
   \end{center}
   \vspace{-0.00cm}
   \caption{#5}
   \label{#6}
  \end{figure}
} 

\begin{document}
%\maketitle

 \title{Differentially Rotating White Dwarfs I: Regimes of Internal Rotation}

\author{Pranab Ghosh\altaffilmark{1,3} and J. Craig Wheeler\altaffilmark{2}}

\altaffiltext{1}{Department of Astronomy \& Astrophysics, Tata 
	Institute of Fundamental Research, Mumbai 400 005, India}

\altaffiltext{2}{Department of Astronomy, University of Texas at
        Austin, Austin, Texas 78712}

\altaffiltext{3}{Current Address: 10/1 Murari Pukur Lane, Kolkata
700 067, India}

\begin{abstract}    

Most viable models of Type Ia supernovae (SN~Ia) require the thermonuclear
explosion of a carbon/oxygen white dwarf that has evolved in a binary system.
Rotation could be an important aspect of any model for SN~Ia, whether single 
or double degenerate, with the white dwarf mass at, below, or above the 
Chandrasekhar limit. {\sl Differential rotation} is specifically invoked in attempts 
to account for the apparent excess mass in the super--Chandrasekhar events. 
Some earlier work has suggested that only uniform rotation is consistent with 
the expected mechanisms of angular momentum transport in white dwarfs, while 
others have found pronounced differential rotation. We show that if the baroclinic 
instability is active in degenerate matter and the effects of magnetic fields are 
neglected, both nearly-uniform and strongly-differential rotation are possible. 
We classify rotation regimes in terms of the Richardson number, Ri. At small 
values of Ri $\leq$ 0.1, we find both the low-viscosity Zahn regime with a 
non-monotonic angular velocity profile and a new differential rotation regime 
for which the viscosity is high and scales linearly with the shear, $\sigma$. 
Employment of Kelvin-Helmholtz viscosity alone yields differential rotation. 
Large values of Ri $\gg$ 1 produce a regime of nearly-uniform rotation for 
which the baroclinic viscosity is of intermediate value and scales as $\sigma^3$. 
We discuss the gap in understanding of the behavior at intermediate values of 
Ri and how observations may constrain the rotation regimes attained by nature.

%240 words, ApJ limit 250

\end{abstract}


%%HERE
\section{Introduction}
\label{intro}

Although the basic explosion mechanism of supernovae of Type Ia
(henceforth SN~Ia) has been established to be the thermonuclear 
combustion of degenerate C/O white dwarfs (henceforth WD),
many aspects of the progenitor systems of remain to be understood. 
Nearly all viable progenitor models involve mass transfer in binary
systems (Howell 2011; Wang \& Han 2102; Maoz, Mannucci \& Nelemans 
2014; but see Chiosi et al. 2014).
One idea is the initiation of carbon ignition as the mass approaches 
the classical Chandrasekhar limit of stability, $\MCh \approx 1.44\Msun$, 
for non--rotating WDs by means of accretion in a binary system. This
classic model is most closely associated with mass transfer from
a non--degenerate companion, the single--degenerate (SD) scenario.
Variations on this theme allow for explosions with less than \MCh\
when accretion of helium from a companion leads to the accumulation 
of an explosive degenerate layer of helium on top of the C/O core 
that generates compression waves that can trigger a central carbon detonation
(Fink et al. 2010; Woosley \& Kasen 2011; Shen \& Moore 2014). This
might occur by accretion from a non--degenerate companion, or from
a degenerate companion in one variety of the double--degenerate (DD)
scenario. Other DD models involve the tidal disruption of one WD,
thus adding mass to the other, resulting in explosion (Dan et al. 2014),
or the violent merger or collision of two WDs (Pakmor et al. 2012; Kushnir
et al. 2013). The DD scenarios typically invoke WDs with mass less than
\MCh, but the total mass might exceed this amount. Other scenarios
manifested in either the SD or DD context involve spinning up a WD
until it is rotationally supported, with a mass exceeding \MCh, 
the explosion being postponed until sufficient angular momentum is lost 
from the star. These are loosely called {\sl spin--up/spin--down} models
(Yoon \& Langer 2005; Di Stefano et al. 2011; Justham 2011; 
Tornambe \& Piersante 2013). In these models, 
the central density rises to the point of carbon ignition
because angular momentum is lost, not because mass is gained.
Yet another variation invokes the merger of a WD with a stellar core
in the context of common--envelope evolution (Livio \& Riess 2003; 
Ilkov \& Soker 2013). 

In the violent merger models, the explosion occurs so quickly
that an issue of the quasi--static rotational state of the WD does not
arise, but in most models the physical context requires the accumulation
of mass and angular momentum, and hence that the WD should rotate.
The rotational state is ignored in many models of the progenitor 
evolution and explosion, including many computationally--demanding
multi--dimensional models; however, it is obvious that, in the
absence of a specific mechanism to lose angular momentum, it must accumulate. 

A generic question that persists throughout the various SN~Ia progenitor 
models is how the internal rotation profile of the accreting WD influences 
the essential dynamical and secular processes that operate in SN~Ia progenitor
environments. One basic influence of the internal rotation has been known 
now for four to five decades, \viz, that (a) uniform rotation can increase the 
maximum mass of stable WDs above the Chandra limit by only about 3 to 4 
percent, and (b) differential rotation can increase this maximum mass substantially, 
exceeding the Chandra limit by factors of up to $\sim 2-3$, and so resulting in 
maximum WD masses up to $\sim (3-4)\Msun$ (see Ostriker \& Bodenheimer 1968 and 
references therein). We shall henceforth refer to \MCh\ as the 
``Chandra limit,'' and WDs with masses above this limit as ``Super--Chandra.''

Most typical SN~Ia are consistent with explosion at \MCh\ (but see Scalzo 
et al. 2014). Polarization observations suggest that most typical SN~Ia may 
not rotate significantly, but that subluminous events (SN~1991bg--like) may 
do so (Patat et al. 2012). Observations of some very bright SN~Ia have led 
to inferred values of the ejecta mass and hence the mass of the pre--SN WD 
in the range (2.1 -- 2.8) \Msun. Examples are SN~2003fg, SN~2006gz, SN~2007if, 
SN~2009dc \citep{howetal,hicketal,yametal,scaletal,silveretal,taubetal}. 
These masses are much in excess of \MCh, and possible implications for 
the internal rotation of the pre--SN WD have been widely discussed. In
this context, one may ask what evolutionary pathway would lead to a WD in a 
slowly--rotating state, and so in a slightly Super--Chandra condition, 
as opposed to a pathway that would lead to a rapidly and differentially
rotating WD in a substantially Super--Chandra condition. If the WD rotates 
rapidly, then issues also arise about the onset of bar--mode instabilities 
that might affect the onset of the explosion and the subsequent dynamics.

In this paper, we focus on clarifying the origins of the different regimes of 
internal rotation that can exist inside accreting WDs. We consider 
WDs without strong, permanent magnetic fields. Accordingly, all angular-momentum 
transport processes considered here are necessarily described by viscosities 
generated only by non--magnetic mechanisms, \eg, hydrodynamic mechanisms 
such as Kelvin--Helmholtz and baroclinic instabilities, and secular mechanisms such 
as the Zahn instability (Zahn 1992). We formulate these viscosities with the aid of 
well--known prescriptions given in the literature. We thus describe angular-momentum 
transport inside accreting WDs in terms of the standard angular--momentum transport 
equation applied in a background WD model, which is pre-specified (\S \ref{elltransport}, 
\S \ref{angmomtransport}). We apply boundary conditions at the WD surface to 
describe the deposition of angular momentum by accreting matter. For low--viscosity 
(Zahn) transport (\S \ref{rothydro}), we recover the differential rotation profile 
found earlier \citep{sainom}. For high--viscosity transport, which includes 
viscosities due to both Kelvin-Helmholtz and baroclinic instabilities, we find both 
(a) the nearly-uniform rotation profile found earlier \citep{sainom,piro}, and, 
(b) a new, differential rotation profile, which is qualitatively 
similar to the inner parts of the differential rotation profile found earlier 
for viscosity due to Kelvin-Helmholtz instability alone \citep{yl04}, but different 
in detail. All of the above regimes of rotation appear to be self-consistent, 
corresponding to different viscosities and different regimes of operation in terms 
of the Richardson number. All regimes described have profiles of specific angular 
momentum (henceforth $\ell$--profile) that increase monotonically outward, 
as they must to be Rayleigh--stable, but the profile of the angular velocity,  
$\Omega$, (henceforth the $\Omega$--profile) can have a maximum inside the WD, 
from which it decreases both outward to the surface and inward to the center.
A general and comparative discussion of these rotation profiles are given in 
\S \ref{rotzahn}, \S \ref{rotyl} and \S \ref{discussregime}. A discussion, our 
conclusions and the prospects for future work are given \S \ref{discuss}.

\section{Angular Momentum Transport Mechanisms}
\label{elltransport}

Mechanisms of angular momentum transport inside rotating stars
can be classified into two general types, \viz, (a) those that
transport angular momentum in the \verthat -direction, 
often called the \emph{vertical} direction, the direction of 
local gravity, and (b) those that transport angular momentum 
in the \horihat -direction, often called the {\sl horizontal} 
direction. The angular momentum is about the stellar
rotation axis of the star, defined as the \zeehat -direction, or 
equivalently the \azihat -direction, in this coordinate system. 
Our concern in this work on accreting WDs is primarily with 
mechanisms of the first type, as these determine the internal rotation 
profile. Mechanisms of the second type work through processes such as 
Eddington-Sweet (henceforth ES) meridional circulation, and the 
Goldreich-Schubert-Fricke (henceforth GSF) instability,  
the primary influence of which is on the distribution of angular 
velocity/momentum in the meridional plane and along the 
\zeehat -axis (\ie, $\Omega (z)$). 

We describe the mechanisms
of vertical or \verthat --transport of angular momentum in this
section, and refer to horizontal or \horihat -transport only 
briefly at appropriate places. Vertical transport mechanisms 
involve turbulent viscosities that can be of two basic kinds:
(a) those that are generated by hydrodynamical instabilities 
that grow on short, dynamical timescales, \eg, the Kelvin-Helmholtz 
instability (henceforth KHI) and the baroclinic instability 
(henceforth BCI), and are generally high, and (b) those that are 
generated by secular instabilities that grow on long, thermal 
timescales, \eg, the Zahn instability (henceforth ZI), and are
generally low. We now discuss these two types of instabilities.    

\subsection{Hydrodynamic Instabilities and Viscous Transport}
\label{hydrovisc}

Among the best--known and most--studied hydrodynamic instabilities 
that grow on dynamical timescales $\sim\Omega^{-1}$ are KHI and 
BCI. The former occurs when there is a (vertical) gradient in 
$\Omega$, and the latter when isobaric and isodensity surfaces 
do not coincide, as is the case when hydrostatic balance has to 
be maintained under differential rotation.   

\subsubsection{Kelvin-Helmholtz instability (KHI)}
\label{KHI}

A vertical gradient in the angular velocity, as measured by the
shear
\begin{equation} 
\sigma \equiv {\partial\Omega\over\partial\ln r},
\label{shear} 
\end{equation}
causes KHI when this shear is sufficiently strong to overcome the 
stabilizing effect of stratification as measured by the 
Brunt-V\"ais\"al\"a frequency, $N$, given by 
\begin{equation}
N^2 \equiv {g\delta\over H_p}\left[\nabla_{\rm{ad}} - 
\left({d\ln T\over d\ln P}\right)\right].
\label{BVfreq}
\end{equation}     
In Equation (\ref{BVfreq}), $\delta\equiv -(\partial\ln\rho/\partial\ln 
T)_P$, $H_p$ is the pressure scale height, and $\nabla_{\rm{ad}}
\equiv(\partial\ln T/\partial\ln P)_{\rm{ad}}$ is the adiabatic 
temperature gradient, the second term in the square brackets being 
the actual temperature gradient at the relevant point inside the 
star. We neglect the effect of composition gradients. The criterion 
for the onset of KHI is expressed in terms of 
the Richardson number Ri, which is defined as
\begin{equation} 
{\rm Ri} \equiv {N^2\over\sigma^2},
\label{richardson} 
\end{equation}
and which compares destabilizing shear with stabilizing buoyancy.
When Ri falls below a critical value, which is widely taken to be
1/4, KHI sets in. We shall adopt this critical value of the 
Richardson number throughout our work here.

KHI leads to a turbulent viscosity, \viscKH, that can be formulated 
\citep{fuji93,sainom,piro} as: 
\begin{equation}
\viscKH = \left\{ \begin{array}{ll}
    {\sqrt{1-4{\rm Ri}}\over 2\sqrt{\rm Ri}}H_p^2N & {\rm Ri} < 1/4 \\
    0 & {\rm Ri} > 1/4.
    \end{array} \right. \;
\label{KHvisc1} 
\end{equation}   
This formulation has been used widely in the literature. An
alternative formulation by \citep{hegetal,yl04} has also been used
widely. It is given by:
\begin{equation}
\viscKH = \left\{ \begin{array}{ll}
    (1-4{\rm Ri})^2H_p^2\Omega_K(r) & {\rm Ri} < 1/4 \\
    0 & {\rm Ri} > 1/4.
    \end{array} \right. \;
\label{KHvisc2} 
\end{equation}   
Here, $\Omega_K(r)\equiv\sqrt{GM_r/r^3}$, is the local Keplerian 
frequency inside the star and, $M_r$ is the mass interior to 
radius $r$. We use both formulations in this work and do not 
find significant differences in the final results.   

\subsubsection{Baroclinic instability (BCI)}
\label{BCI}

Surfaces of constant pressure and density may no longer coincide 
when hydrostatic equilibrium is maintained under differential
rotation. Their misalignment causes BCI, the characteristics of which
depend on both the Richardson number, Ri, introduced above, and
a critical baroclinic Richardson number, Ri$_{\rm BC}$, given by
\begin{equation} 
{\rm Ri}_{\rm BC} \equiv 4\left({r\over H_p}\right)^2{\Omega^2\over N^2},
\label{richBC} 
\end{equation}  
above which Coriolis forces limit the scale of the horizontal 
perturbations and so reduce the strength of the BCI somewhat.
  
BCI leads to a turbulent viscosity, \viscBC, which has been given
\citep{fuji93,sainom,piro} as:
\begin{equation}
\viscBC = \left\{ \begin{array}{ll}
    {1\over 3{\rm Ri}^{1/2}}H_p^2\Omega & {\rm Ri} < {\rm Ri}_{\rm BC} \\
    {{\rm Ri}_{\rm BC}\over 3{\rm Ri}^{3/2}}H_p^2\Omega 
       & {\rm Ri} \gg {\rm Ri}_{\rm BC}.
    \end{array} \right. \;
\label{BCvisc} 
\end{equation}
We use the prescription of Equation (\ref{BCvisc}) in this work.

Angular momentum transport by BCI-generated viscosity for accreting 
degenerate stars has been studied by Fujimoto (1993), and detailed results 
for accreting WDs with BC viscosity included have been obtained by Saio \&
Nomoto (2004) and by Piro (2008). The work of Yoon \& Langer (2004) is
relevant in this context for having excluded BCI. We have included here both 
KH and BC viscosities (\ie, \viscKH\ and \viscBC), in order to have a 
complete discussion of all possible rotation regimes found so far.
The rotation regime found by Yoon \& Langer (2004), who included 
only the KH viscosity among the hydrodynamic ones, together with 
the secular Zahn viscosity and processes like the ES circulation and 
the GSF instability, is described in \S \ref{rotyl}. 
The inclusion of the BCI yields the result that there are two
asymptotic, disparate rotation regimes.

Although the overall prescription for the baroclinic viscosity is customarily 
given a single label BC, we see from Equation (\ref{BCvisc}) that there are
two different asymptotic regimes depending on the Richardson number, Ri.
The actual viscosities are quite \emph{different} in the two regimes, 
with different scalings. According to Equation (\ref{BCvisc}), the BC viscosity, 
\viscBC, scales with the Richardson number Ri as ${\rm Ri}^{-1/2}$ in the
regime of small Ri, while it scales as ${\rm Ri}^{-3/2}$ in the regime of 
large Ri. Since Ri scales with the shear as Ri $\propto\sigma^{-2}$, we see 
that \viscBC\ scales as $\sigma$ in the small Ri regime, but as $\sigma^3$ 
in the regime of large Ri. The addition of the KH viscosity to 
the BC viscosity does not change these scalings significantly, since \viscKH\ 
also scales with Ri roughly as ${\rm Ri}^{-1/2}$ (Equation (\ref{KHvisc1})) in 
the small Ri regime while \viscKH\ vanishes in the regime of high Ri-values.    
Omission of the BC viscosity yields a regime dominated by the KH viscosity. 
This regime is roughly comparable to the small Ri BC regime and corresponds 
only to differential rotation.

From these considerations, we see that in the presence of the BCI the 
viscous stress, $\tau = \nu\sigma$, scales \emph{nonlinearly} with the shear in 
the regimes of both small and large Ri, with $\tau \propto \sigma^2$ in the 
small Ri regime and $\tau \propto \sigma^4$ in the large Ri regime. This 
difference leads to a duality in the rotation profiles, essentially solid body and 
substantially differential, even with the same initial and boundary conditions, 
depending on the effective value of Ri that in turn depends on the shear in the 
self-consistent solution. 

In practice, these two different regimes were reproduced by implementing 
Equation (\ref{BCvisc}) in different, but formally equivalent, ways in the two 
different regimes. One set of solutions began with no rotation, a given set of 
boundary conditions and Ri $<<$ 1. This formulation evolved (in $MATLAB$) 
a differential rotation solution, new to this work, in which both the KH and 
BC viscosities contributed. This solution remained in the small Ri regime, 
specifically Ri $\sim 10^{-2}$, where the viscosity scales as $\sigma$
and $\tau$ scales as $\sigma^2$ (\S\ref{diff}). The second formulation again 
adopted Equation (\ref{BCvisc}), but recast in terms of the variable 1/Ri, which 
is a small quantity in the regime of large Ri. Solving the same equations with 
the same boundary conditions but beginning the evolution with 1/Ri $<<$ 1 
evolved a solution that, after short transient phase, attained and remained 
in a state of nearly solid-body rotation (\S\ref{uniform}) in the regime of large Ri, 
Ri $\sim 10^6$, with viscosity scaling as $\sigma^3$, and $\tau$ scaling as 
$\sigma^4$. This solution was dominated by the BC viscosity and had negligible 
contribution from the KH viscosity. This procedure is sufficient to establish that 
Equation (\ref{BCvisc}) supports at least two very different, self-consistent, rotation 
states. 

The question of how the solutions evolve for the large range of intermediate 
values of Ri with yet different scalings with Ri and $\sigma$ remains open. 
Omitting BCI, the formulation of Yoon \& Langer (2004) is dominated by the KH 
viscosity in the inner regions and corresponds to Ri very slightly below 0.25 in 
the inner regions of their models, near the low Ri regime as we define it here 
(\S\ref{rotyl}). Uniqueness theorems say that the solution in each power-law limit, 
$\viscBC \propto \sigma$ and $\viscBC \propto \sigma^3$ is unique but do not 
constrain the behavior in intermediate regimes. We return to a discussion of these 
multiple solutions in \S\ref{discussregime} and \S\ref{discuss}. 

\subsection{Secular Instability and Viscous Transport}
\label{secvisc}

Even when stable stratification prevents dynamical instabilities, 
secular instabilities are still possible if there is sufficiently 
strong thermal diffusion that reduces the buoyancy force through 
radiative leakage. This phenomenon is described in terms of the 
P\'eclet number, Pe, which in this context is essentially the ratio 
of the rate of advection of momentum by the turbulent flow to the 
rate of thermal diffusion. The  P\'eclet number is given for this 
particular problem by ${\rm Pe}\equiv vl/K$, where $v$ and $l$ are 
respectively the eddy velocity and the size of turbulent eddies, 
and $K$ is the thermal diffusivity. Thermal diffusion modifies the 
criterion for the onset of instability from Ri $<$ 1/4 to 
Ri $\times$ Pe $<$ 1/4.

The resultant secular instability, \ie, the Zahn instability (ZI)
leads to a turbulent viscosity, \viscZ, that was calculated by 
Zahn (1992), and can be expressed as:
\begin{equation}
\viscZ = {2\over 45}{K\over {\rm Ri}}.
\label{Zvisc}
\end{equation}
Here, $K\equiv 4acT^3/(3\kappa\rho^2C_p)$ is the thermal diffusivity.
We use the above prescription for ZI viscosity in this work.
           
\section{Angular Momentum Transport Inside White Dwarfs}
\label{angmomtransport}

In this work, we describe angular momentum transport inside 
accreting WDs by solving the transport equation in the background
of a WD model with specified structure, thus neglecting the 
effect of changing rotation on WD structure.
We adopt this ``toy'' approach here for the purpose of clarifying 
the possible basic regimes of rotation profiles in accreting WDs. 
We do emphasize at this point that the effects of accretion
are explicitly taken into account in our work by (a) keeping track of 
the changes in the mass and radius of the white dwarf as a result of
accretion, as described below, and, (b) representing the inward
transport of angular momentum by the inward motion of mass-shells 
in an accreting, contracting WD by a suitable ``advection'' term,
as explained below and detailed in Appendix A.   

The equation for angular momentum transport inside the WD can be 
expressed as 
\begin{equation}
{\partial\over\partial t}\left[r^2\Omega\right] = {1\over 4\pi\rho
  r^2}{\partial\Sigma\over\partial r}.
\label{transport1} 
\end{equation}
Here, $\Sigma(r,t)$ is the total rate of angular 
momentum transport, through radius $r$ at time $t$, \ie, the total 
torque at that radius and time. If this transport were entirely 
viscous, the rate would simply be $\Sigma = 4\pi\rho\nu r^4\partial
\Omega/\partial r$; however, accreting WDs contract in response to
their increase in mass. Consequently, each mass shell moves inward, 
carrying angular momentum with it. This behavior constitutes, in effect,
a slow, inward advection of angular momentum through the WD on the
accretion timescale \citep{yl04}. This advection is automatically 
taken into account if one recalculates the WD model at each step
as mass and angular momentum accretion proceeds. Since we 
are working here with a backgound WD structure unaffected by changing
rotation, we have to account for this advection effect explicitly. This 
is easily done, as we do take into account the contraction of the WD
as it accretes mass, as described below.
We have chosen the term ``advection'' for this effect to stress that
this part of the transport is actually due to the slow inward motion of 
mass shells (in Eulerian co-ordinates) as an accreting white dwarf 
adjusts its internal structure, and not due to viscosity. Alternatively, 
if we visualize the situation in terms of Lagrangian co-ordinates,  
we can look upon this effect as being due to local conservation of 
angular momentum. We emphasize that if we define a formal advection 
velocity $v_r$ (see below) for this effect, it will be much smaller 
than the velocities that occur, \eg, in advective accretion disks.  

We now model the above effect in a straightforward manner in terms of 
a very slow, inward flux of matter through the WD at a rate 
$\Mdot(r,t)$, which depends on the Eulerian radius $r$ and the time
$t$. We note first that we can express this slow advection as 
$\Mdot(r,t) = 4\pi r^2\rho(r,t)v_r(r,t)$ 
to define a formal advection velocity $v_r(r,t)$,
which is actually determined by the slow contraction of the WD as its
mass increases due to accretion, expressed through the mass-radius
relation $\RWD = \RWD(\MWD)$, with \RWD\ decreasing as \MWD\ increases.

Power-law prescriptions, $\RWD\sim\MWD^{-s}$, have been widely used
as approximations to numerical results for mass-radius relations 
obtained from detailed models. For low-mass WDs, the non-relativistic
result $s = 1/3$ holds well, while the power steepens as the Chandra
limit is approached. As necessary, we have considered in the rest of
this work both the non-relativistic limit of $s$ and $s$-values
$\sim 1$ inferred from numerical WD models in the literature 
\citep{yl04,sainom}. 

To proceed further, we require a prescription for the 
contraction of the interior mass-shells as the whole WD contracts. We 
describe our prescription in Appendix A, where we show that 
the advection rate, $\Mdot(r,t)$, can be reasonably approximated 
by a relation of the form
\begin{equation}
\Mdot(r,t) \approx M_r{\Mdot_{WD}\over\MWD},
\label{advrate}
\end{equation}  
where $\Mdot_{WD}$ is the mass accretion rate onto the WD. With this
term included, the complete expression for the angular momentum
transport rate, or the total torque, becomes
\begin{equation}
\Sigma = 4\pi\rho\nu r^4{\partial\Omega\over\partial r} + 
\Mdot(r,t)r^2\Omega,
\label{totrate}
\end{equation}    
with \Mdot(r,t) given by Equation(\ref{advrate}).

The equation for angular momentum transport inside the WD is obtained 
by combining Equations (\ref{transport1}), (\ref{advrate}), and 
(\ref{totrate}). We then change the dependent variable to the specific
angular momentum, $\ell\equiv\Omega r^2$, and the independent variable 
to the mass co-ordinate, $M_r$,  which is related to
the radial co-ordinate, $r$, through $dM_r = 4\pi\rho r^2dr$. The final
result is:  
\begin{equation}
{\partial\ell\over\partial t} = {\partial\over\partial M_r}\left[
4\pi\rho r\nu\left(4\pi\rho r^3{\partial\ell\over M_r} - 2\ell\right)
+ {M_r\ell\over t_{ac}}\right].
\label{transport2} 
\end{equation}
Here, $t_{ac}\equiv\MWD/\Mdot_{WD}$ is the accretion timescale. 
          
We solve Equation (\ref{transport2}) numerically for the various types 
of viscosities described above. For the background WD structure
in which this transport occurs, we adopt the numerical WD models
supplied to us by Montgomery (2010). The initial model, on which
accretion begins, is a C/O-core WD described by the following 
parameters: a mass of $\MWD = 1.2\Msun$, an effective surface gravity 
(in cgs units) of $\log g = 9.022$, and an effective surface 
temperature of $T_{eff} = 1.2\times 10^4$ K. The model has an 
appropriate gradient in the C/O ratio in the core, and there are
layers of He and H on top of this core. As accretion proceeds, the
WD contracts according to scheme described above, which is reflected
in the outer boundary condition indicated below.   
      
For our numerical solutions, we need appropriate initial conditions.
We also need appropriate boundary conditions at the center and surface of the 
accreting WD. For the former, we have taken WDs that are nonrotating 
initially. For the latter, consider the center of the star first, 
where the boundary condition we apply is $\ell_{cent} = 0$ in all cases, 
remembering that the transport equation (Equation (\ref{transport2})) is 
formulated in terms of $\ell$. We emphasize here
that the angular velocity at the center, $\Omega_{cent}$, is not 
directly constrained in our solutions. Rather, it evolves 
along with the solution throughout the star according to the 
particular prescription for the viscosity. Now consider the surface 
of the star, where we have explored the effects of applying the two 
types of boundary condition suggested previously in the literature. 
These are: (1) the boundary condition of Saio \& Nomoto (2004),
(henceforth referred to as the {\sl SN boundary condition}), wherein the 
angular velocity at the surface of the WD is set to the {\sl current} 
Keplerian value $\Omega_K(\RWD)$ there, and, (2) the boundary condition 
of Yoon \& Langer (2004), (henceforth referred to as the {\sl YL 
boundary condition}), wherein each element of accreted mass 
$\Delta M$ deposits an amount of angular momentum on the WD that is 
(a) $\Delta M\ell_K(\RWD)$ if the surface angular velocity 
$\Omega(\RWD)$ of the WD is less than $\Omega_K(\RWD)$, and
(b) zero if $\Omega(\RWD)\ge\Omega_K(\RWD)$. Here, $\ell_K(\RWD)\equiv
\Omega(\RWD)\RWD^2$ is the specific angular momentum in a Keplerian
orbit at \RWD. 
We emphasize that quantities such as $\ell_K(\RWD)$ 
and $\Omega_K(\RWD)$ are understood here to be the {\sl current} 
values, corresponding to the current values of $\MWD$ and $\RWD$
during the process of accretion.

Although the SN and YL boundary conditions work with the same or 
closely related variables, their stipulations are not exactly
the same.
We explore the effects of SN and YL boundary conditions in 
our work, using a simple version of the latter condition adequate
for our purposes, which we describe in Appendix B. We find only
minor differences in the final outcome. Where the differences are
negligible, we quote only one result.

\section{Regimes of Internal Rotation for Hydrodynamic Viscosities}
\label{rothydro}

We now present the internal rotation profiles for the viscosities
generated by hydrodynamic instabilities, \ie, KHI and BCI. With both 
KH and BC viscosities included in our calculations, we find two 
regimes of rotation as sketched in \S 2.1.2, \viz, (a) nearly-uniform 
rotation, and, (b) strongly differential rotation. The former regime has 
been discussed earlier \citep{sainom,piro}. We consider each regime in turn.
 
\subsection{Nearly-uniform Rotation}
\label{uniform}

The regime of nearly-uniform rotation corresponds to the asymptotic
regime of large Ri
(Ri $\sim 10^6$ in our solutions),
 $\viscBC \propto \sigma^3$ 
and $\viscKH = 0$ described in \S\ref{BCI}. The solution is illustrated in Figure 
\ref{uniformOmega}. The left panel is a 3-D surface plot of the evolution 
of the $\Omega$--profile as accretion proceeds, while the right panel 
shows this evolution as stacked profiles in a 2-D plot as time elapses, 
the topmost curve giving the final time-step. Note that the mass
co-ordinate for the profiles in this figure, and all subsequent 
evolutionary figures in this work, is $M_r/\MWD$, with $M_r$ as 
defined earlier, and $\MWD$ the current mass of the WD. Also note 
that time axis is marked in this figure in units of the accretion 
timescale as defined earlier, which is much longer than the viscous 
timescale for the high KH and BC viscosities (see \S\ref{discussregime}). 
Evolution of the $\Omega$-profile in this case is basically the increase 
of this nearly-uniform $\Omega$-value as accretion proceeds and the WD 
contracts. 

In practice, our solutions in this regime start with no rotation. As 
soon as angular momentum is added by means of a boundary condition, 
strong gradients, and hence finite viscosities, are temporarily generated in 
the outer layers. The initially large viscosity quickly leads to nearly solid-body 
rotation with lower viscosity and slightly, but monotonically, increasing angular 
velocity with radius.  The viscosity in these solutions is $\nu_{BC} \sim 10^8$ 
cm$^2$s$^{-1}$, which we label an {\sl intermediate viscosity},  in contrast with 
other solutions described below. We emphasize that, for this regime as well 
as that of strong differential rotation described in the next subsection, the
angular momentum transport equation was always solved self-consistently by 
evaluating Equation 7 according to the regime of locally large or small Ri at each location at a given time. We also note that the ``simple" solid-body solution 
described in this subsection is not possible in a formulation that neglects the 
BCI and hence is dominated by the KHI.

 \insertdoblfig{scale=0.5}{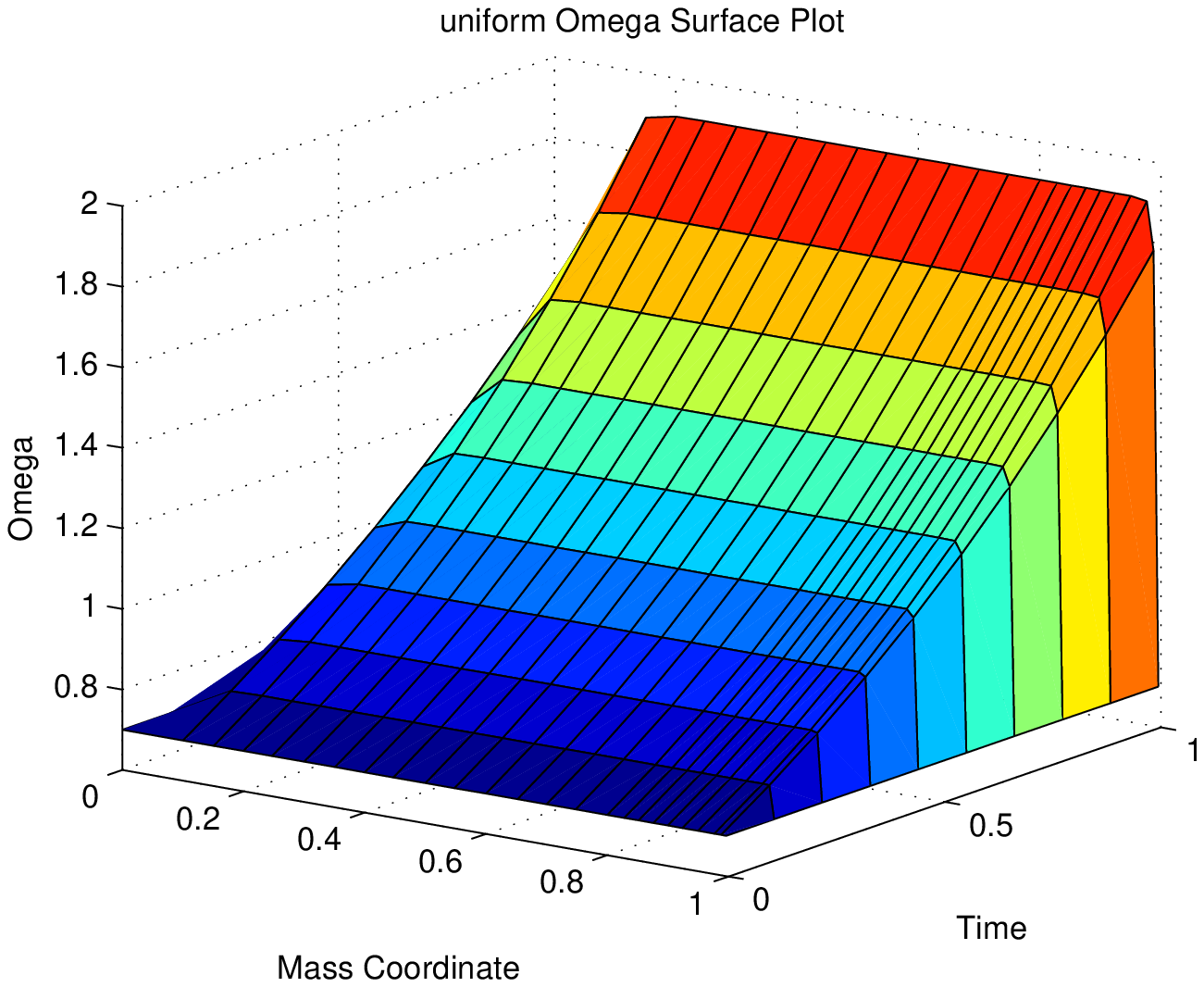}
             {scale=0.5}{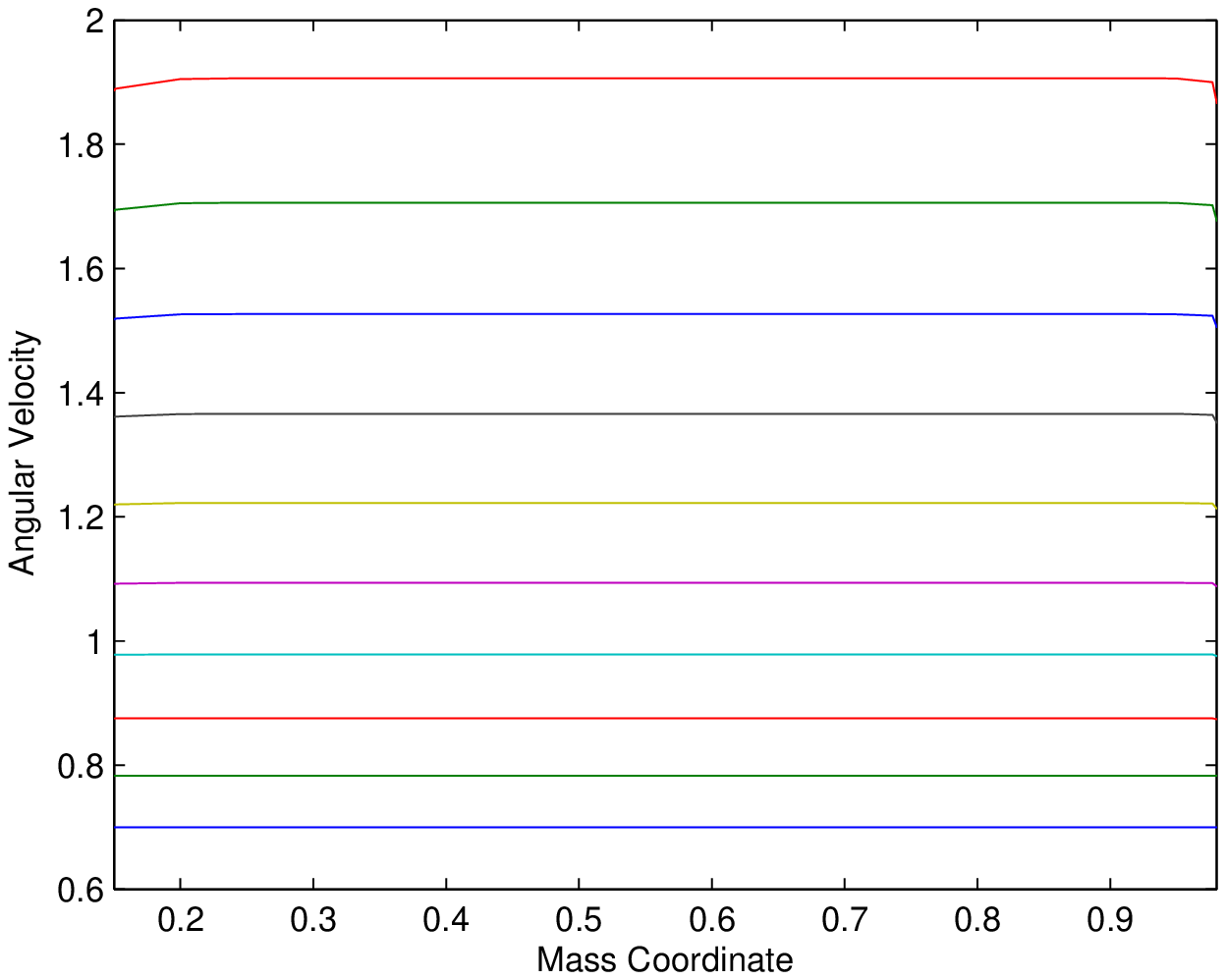}
{\it Evolution of the angular velocity profile
(in units of s$^{-1}$) in the nearly-uniform rotation case 
with KH and BC viscosities. The time axis 
is marked in units of the accretion timescale. Left Panel: 3-D 
surface plot. Right Panel: Stacked profiles in a 2-D plot, wherein 
the profiles at various time-steps from the left panel are 
color coded, with time going up vertically.}{uniformOmega}   

\subsection{Strongly Differential Rotation}
\label{diff}

We find that strong differential rotation is also possible when viscous 
transport includes both KH and BC viscosities or either viscosity 
prescription alone.The regime of strong differential rotation corresponds 
to the asymptotic regime of small Ri where both \viscKH\ and \viscBC\ 
scale approximately proportionally to $\sigma$ as described in \S\ref{BCI}.
We find the effective viscosity in this regime to be $10^{15}$ to
$10^{16}$ cm$^2$s$^{-1}$ (see Figure \ref{GWdetail} below) that we
label a regime of {\sl high viscosity}.
A typical example is shown in detail in Figures \ref{GWOmega1},
\ref{GWOmega2}, \ref{GWOmega3}, and \ref{GWOmega4}.
In each figure, the left panel displays in a 3-D surface plot the 
evolution of the $\Omega$--profile as accretion proceeds. {\it 
The time axis is marked in these figures in units of the viscous timescale, 
in order to study how the profile ``heals'' to an asymptotic form 
on this timescale, and how that form evolves as accretion proceeds
on a much longer timescale.} The right panel of the same figure shows 
the same evolution as stacked profiles in a 2-D plot as time elapses, 
the topmost curve giving the final time-step.

We first explore the effect of SN boundary condition. Figure 
\ref{GWOmega1} shows the approach to the asymptotic state,
wherein we follow the evolution to a maximum time $t_{max} =
4$ in units of the viscous time. On the 2D plot
in the right panel, the results are displayed in 10 equal steps
over the time span 0 to $t_{max}$, the topmost curve corresponding
to $t_{max}$. 

\insertdoblfig{scale=0.5}{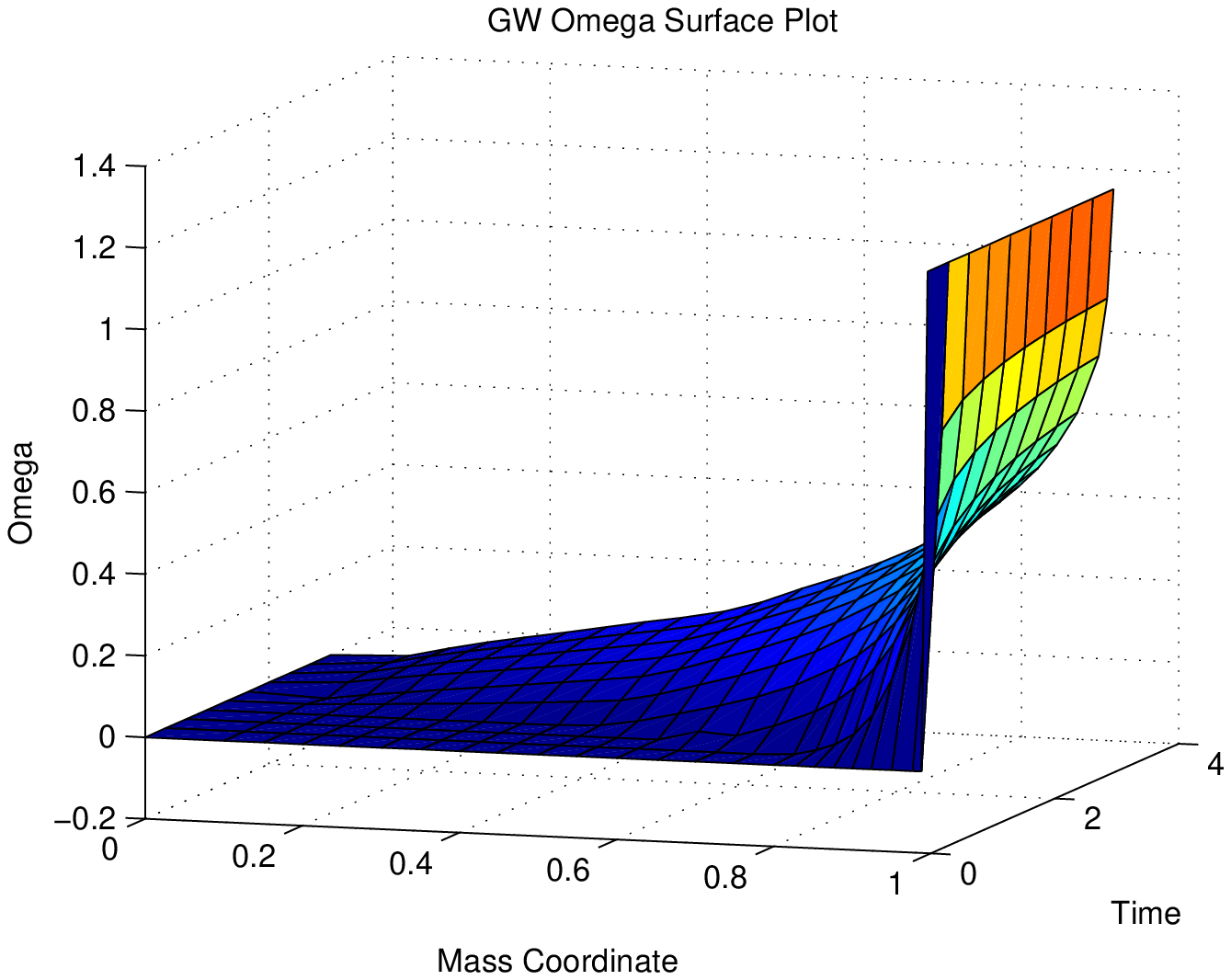}
             {scale=0.5}{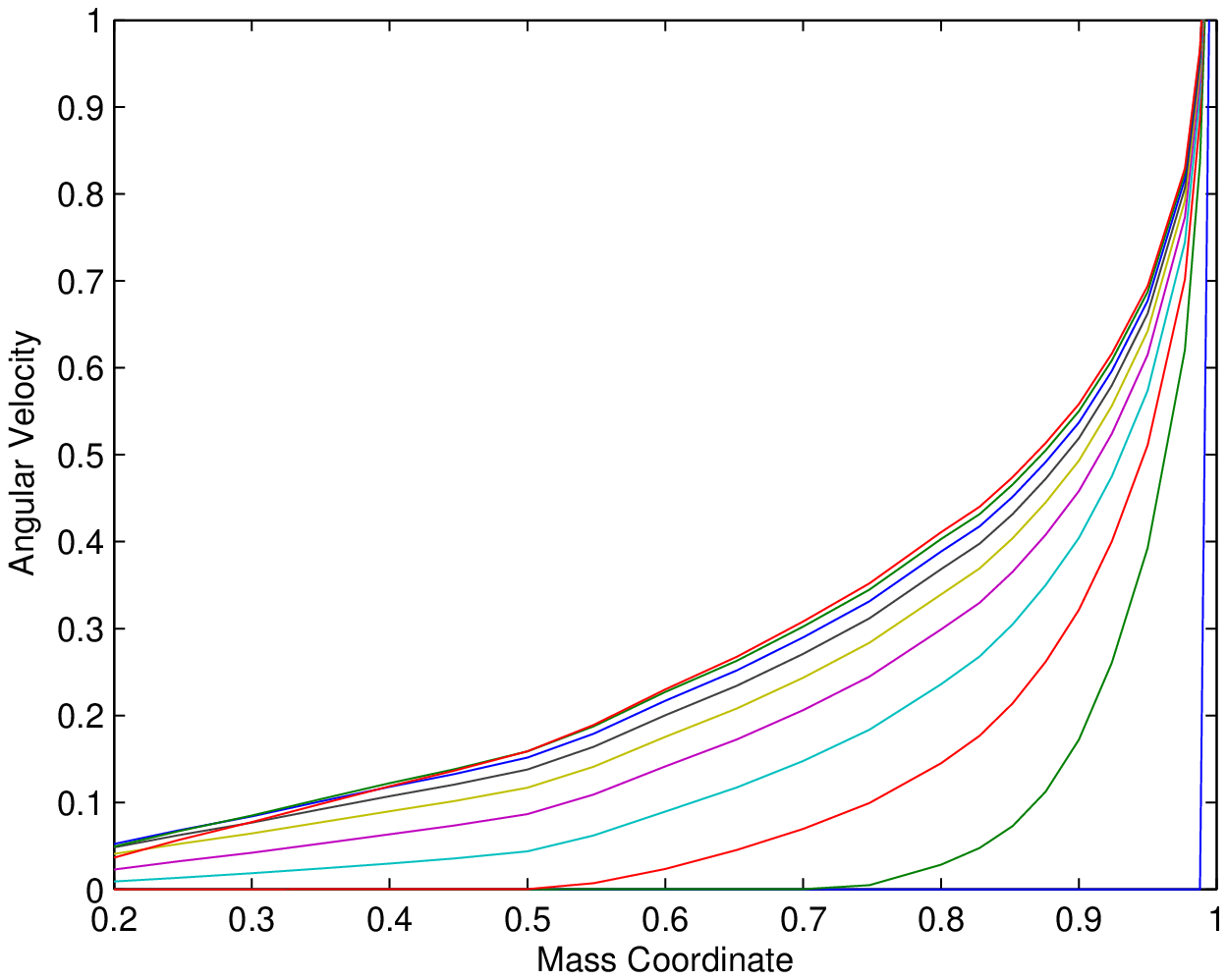}
{\it Evolution of the angular velocity profile
(in units of s$^{-1}$) in the differential 
rotation case with KH and BC viscosities, and SN boundary condition
(see text). The time axis is marked in units of the viscous timescale, 
the maximum time $t_{max}$ being 4 in these units. Left Panel: 3-D 
surface plot. Right Panel: Stacked profiles in a 2-D plot, wherein 
the profiles at 10 equal time-steps are color coded, with time going 
up vertically.}{GWOmega1} 

We show longer evolution of this profile in Figure 
\ref{GWOmega2}, where we increase $t_{max}$ to 500 in the same
units. On the 2D plot in the right panel, the results are displayed 
in 10 equal steps over the time span 0 to $t_{max}$, the topmost 
curve corresponding to $t_{max}$. We see that the approach to the
asymptotic profile is rapid, attaining it practically within the
first time-step. Hence the profiles are almost indistinguishable
from one another in the 2D plot.

\insertdoblfig{scale=0.5}{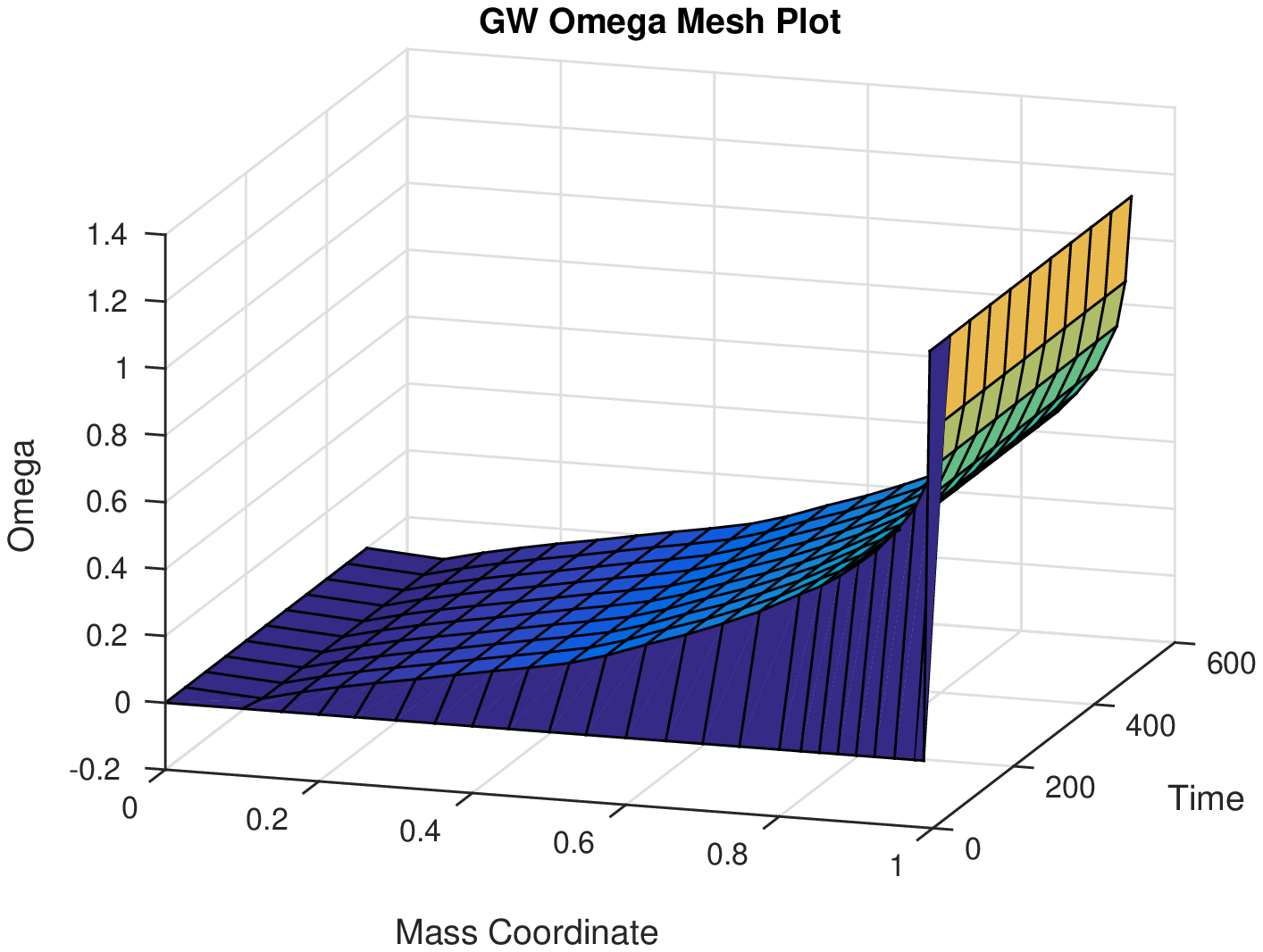}
             {scale=0.5}{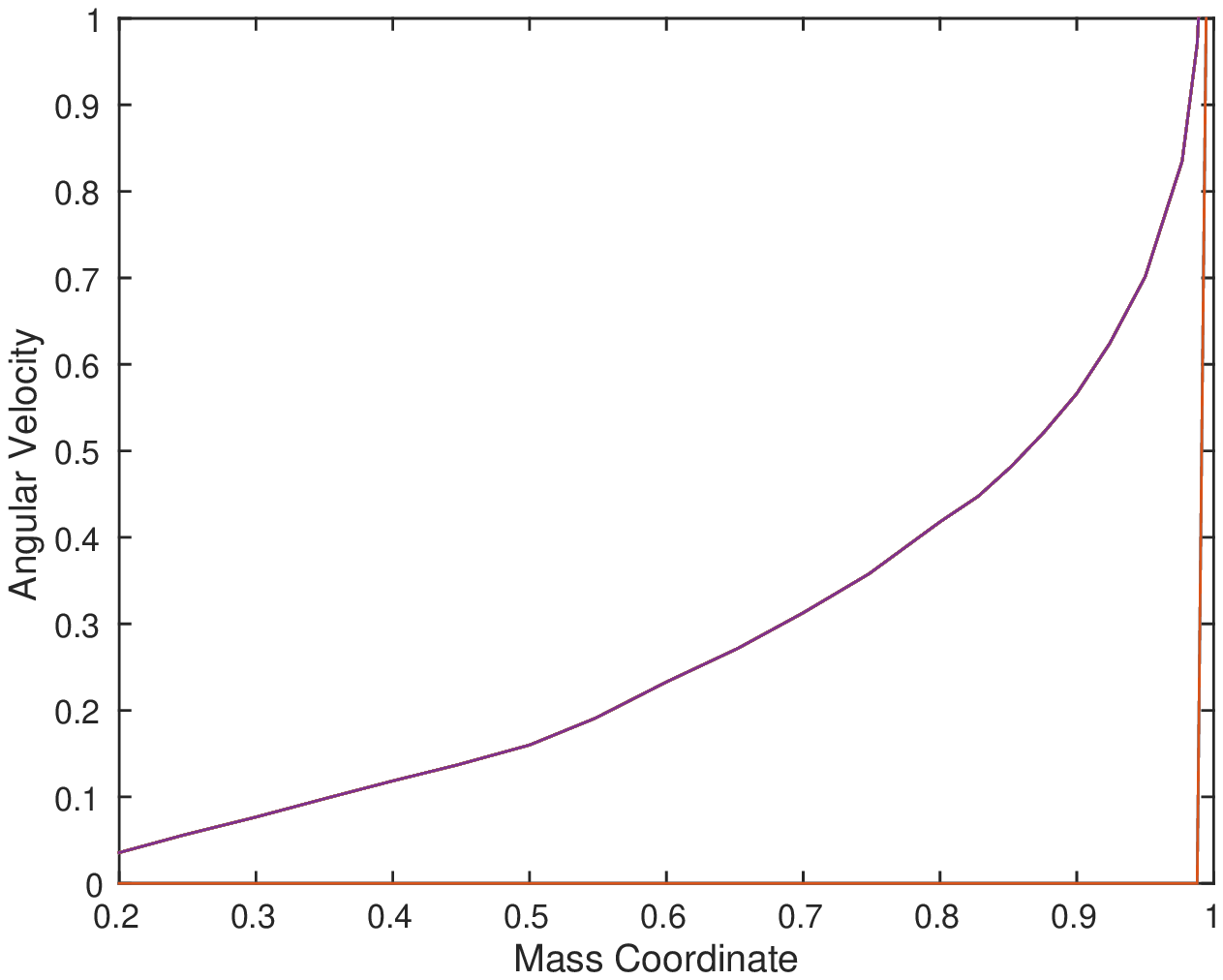}
{\it Evolution of the angular velocity profile
(in units of s$^{-1}$) in the differential 
rotation case with KH and BC viscosities, and SN boundary condition
(see text). The time axis is marked in units of the viscous timescale, 
the maximum time $t_{max}$ being 500 in these units. Left Panel: 
3-D surface plot. Right Panel: Stacked profiles in a 2-D plot, 
wherein the profiles at 10 equal time-steps are color coded, with 
time going up vertically.}{GWOmega2} 

We show even longer evolution of this profile in Figure 
\ref{GWOmega3}, where we increase $t_{max}$ to 50,000 in the same
units to confirm the result. We do not repeat the 2D plot in this
case, as the rapid attainment of the asymptotic profile makes the 
10 equal time-step profiles completely indistinguishable from one 
another on this plot. 

\insertfig{scale=0.5}{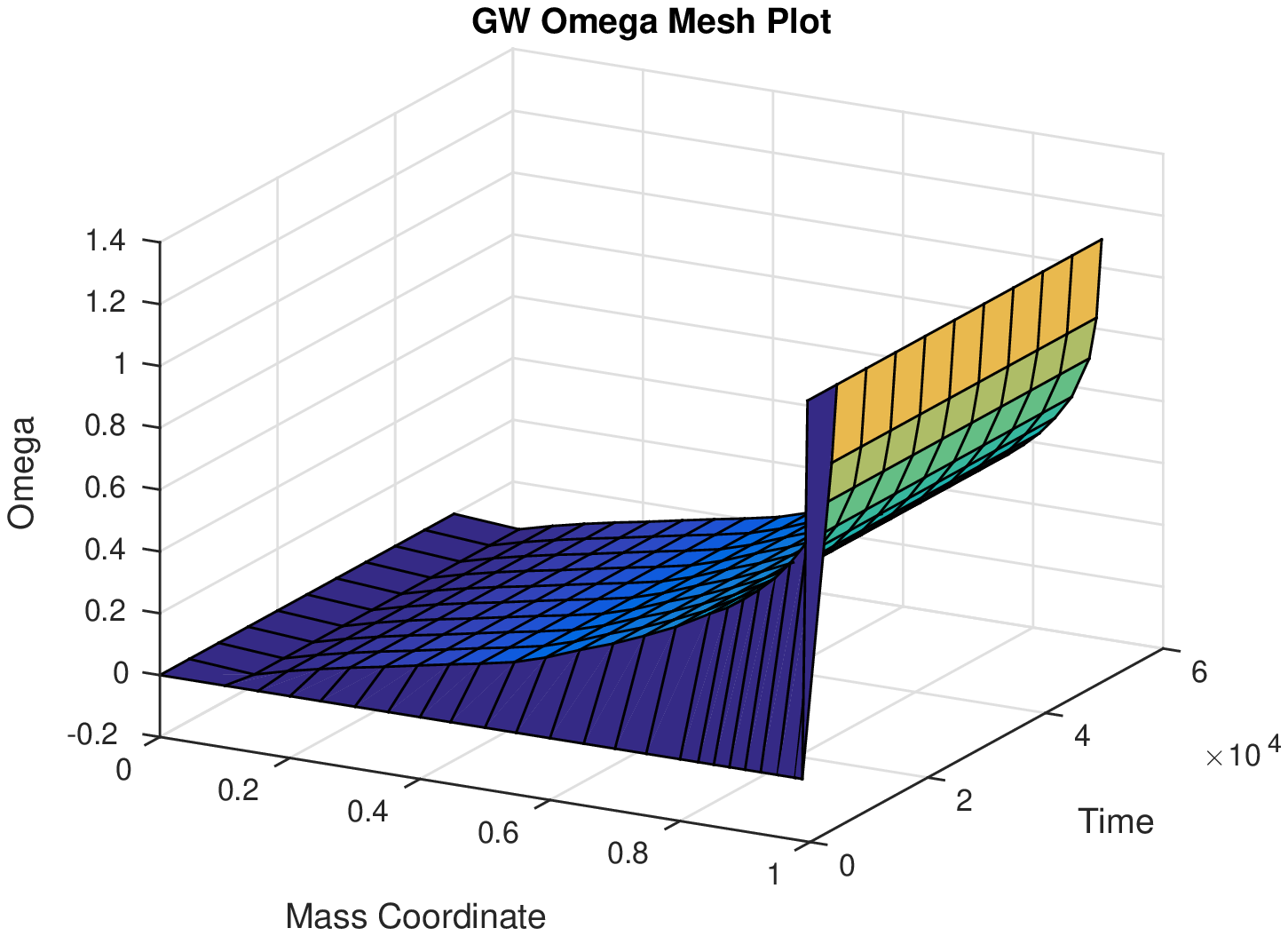}
%%             {scale=0.5}{2DSN5e4.eps}
{\it Evolution of the angular velocity profile 
(in units of s$^{-1}$) in the differential 
rotation case with KH and BC viscosities, and SN boundary condition
(see text). The time axis is marked in units of the viscous timescale, 
the maximum time $t_{max}$ being 50,000 in these units. Only the 
3-D surface plot is shown in this case, for reasons explained in the
text.}{GWOmega3} 

We now explore the effect of the YL boundary condition. We mimic 
this boundary condition by a simple prescription detailed in Appendix B,
adequate for our purposes. Basically, as angular momentum is added
to a non-rotating WD, the specific angular momentum at its surface,
$\ell_{surf}$, increases according to the prescription
\begin{equation}
{\partial\ell_{surf}\over\partial t} = {\ell_K(\RWD)\over t_{ac}},
\label{YLbc}
\end{equation}  
until the the angular velocity at the surface reaches the Keplerian
value there. At that point, the surface angular velocity is 
maintained at the Keplerian angular velocity (which changes with
time as accretion continues as the WD contracts). 
In Equation (\ref{YLbc}), $\ell_K(\RWD) = \sqrt{G\MWD\RWD}$ is the 
Keplerian value of $\ell$ at the surface of the WD of mass \MWD\
and radius \RWD . In reality, after the surface angular velocity 
reaches the Keplerian value for the first time, the value of the 
Keplerian angular velocity at the surface increases as the WD 
contracts due to accretion, so that the actual surface angular 
velocity becomes sub-Keplerian, and so angular momentum is added 
to it by the accreting mass, spinning it to the Keplerian 
value again, and so on \citep{yl04}. This prescription is an 
adequate approximation for our purposes here.  

In Figure \ref{GWOmega4}, we show the evolution of the rotation 
profile with YL boundary condition and with a WD mass-radius 
relation described by a power-law index $s=1.2$ (see above), which 
roughly describes many of the computed WD models of Yoon and Langer 
(2004; see their Table 2). We follow the evolution 
to a maximum time $t_{max} = 5000$ in the same units. We 
model the approach to Keplerian rotation at the surface by 
assuming that it is attained in a time $t_{K} = 0.5t_{max}$ 
and maintained at Keplerian thereafter. As shown in Appendix B,
$t_{K} = x_c t_{ac} \approx t_{ac}$, so that our choice corresponds
to $t_{ac} \approx 0.5t_{max} = 2500 t_{visc}$ in this example.
See further discussion below.

As Figure \ref{GWOmega4} shows, the asymptotic profile is 
obtained soon after Keplerian rotation is reached at the surface 
and changes little on further evolution.       

\insertdoblfig{scale=0.5}{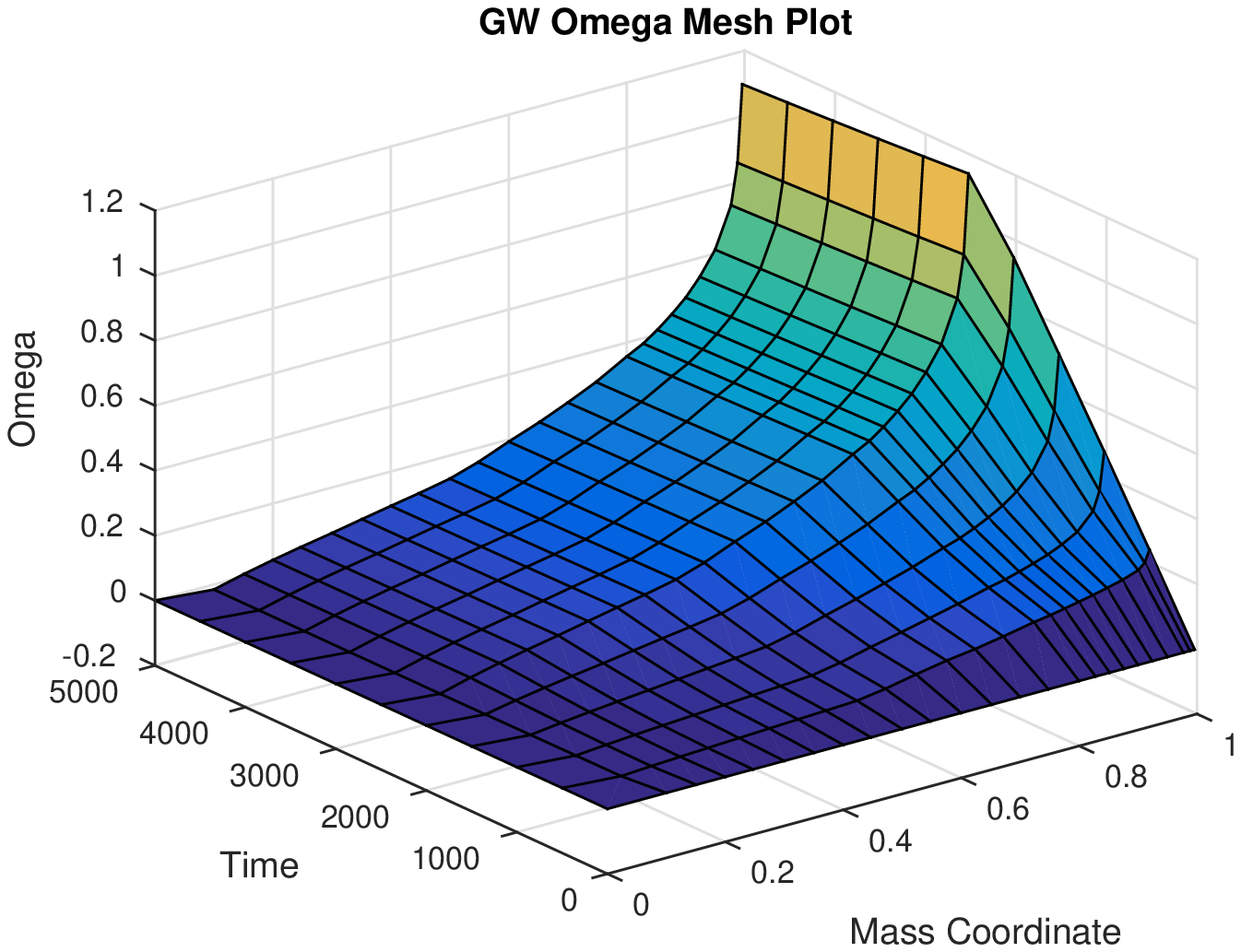} 
             {scale=0.5}{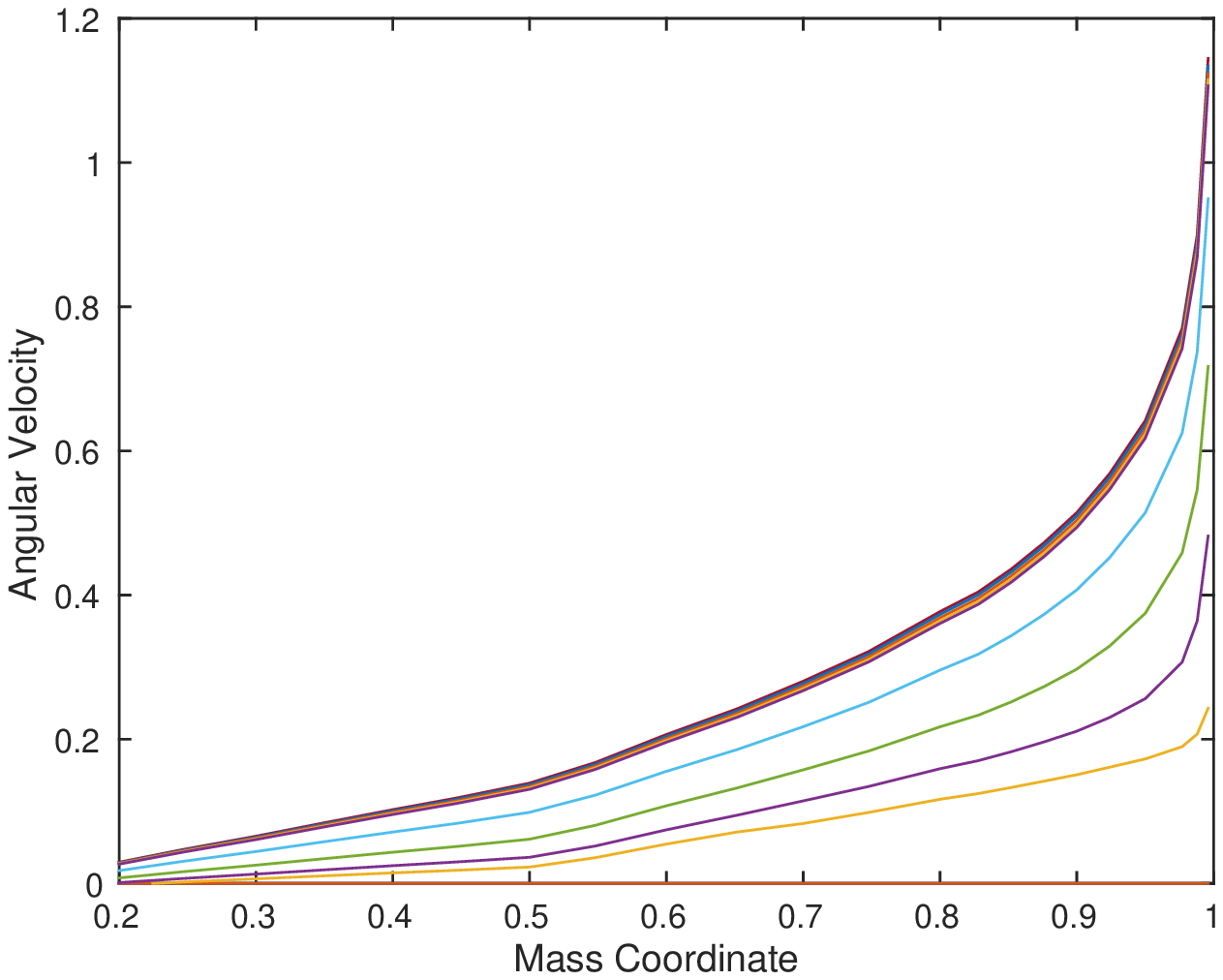}
{\it Evolution of the angular velocity profile (in units of s$^{-1}$)
 in the differential 
rotation case with KH and BC viscosities, and YL boundary condition,
with a mass-radius relation with power-law index $s = 1.2$ (see text). 
The time axis is marked in units of the viscous timescale, 
the maximum time $t_{max}$ being 5000 in these units. Left Panel: 
3-D surface plot. Right Panel: Stacked profiles in a 2-D plot, 
wherein the profiles at 10 equal time-steps are color coded, with 
time going up vertically.}{GWOmega4} 

The evolutions shown in Figures \ref{GWOmega1}, \ref{GWOmega2}, 
\ref{GWOmega3}, and \ref{GWOmega4} clarify the nature of the 
asymptotic rotation profile and demonstrate that SN or YL boundary
conditions make little difference to the final profile. We stress at
this point that we have used rather large values of $t_{max} \sim 10^3 
- 10^4$ in some of these calculations in order to simulate the effect 
of an accretion timescale $t_{ac}$ which is much longer than the 
viscous timescale $t_{visc}$ and demonstrate how the asymptotic 
profile is approached. For the large hydrodynamic viscosities 
generated by KHI and BCI, the actual value of $t_{ac}/t_{visc}$ is 
much larger (see \S \ref{discussregime}) and is not possible in 
practice to achieve computationally; however, the asymptotic approach 
shown above makes this unnecessary. 

The profiles of the viscosities \viscKH\ and \viscBC\ (in cgs units) 
are shown in the left panel of Figure \ref{GWdetail} for the
asymptotic profile, calculated self-consistently during the
computation. The viscosity is itself dependent on the 
$\Omega$--profile through the shear and hence the Richardson number 
(see \S \ref{hydrovisc}). Shown in the right panel of the same 
figure are the profiles of the Richardson number Ri and its 
critical value ${\rm Ri}_{\rm BC}$ defined in \S \ref{hydrovisc}, 
for this asymptotic profile.

\insertdoblfig{scale=0.5}{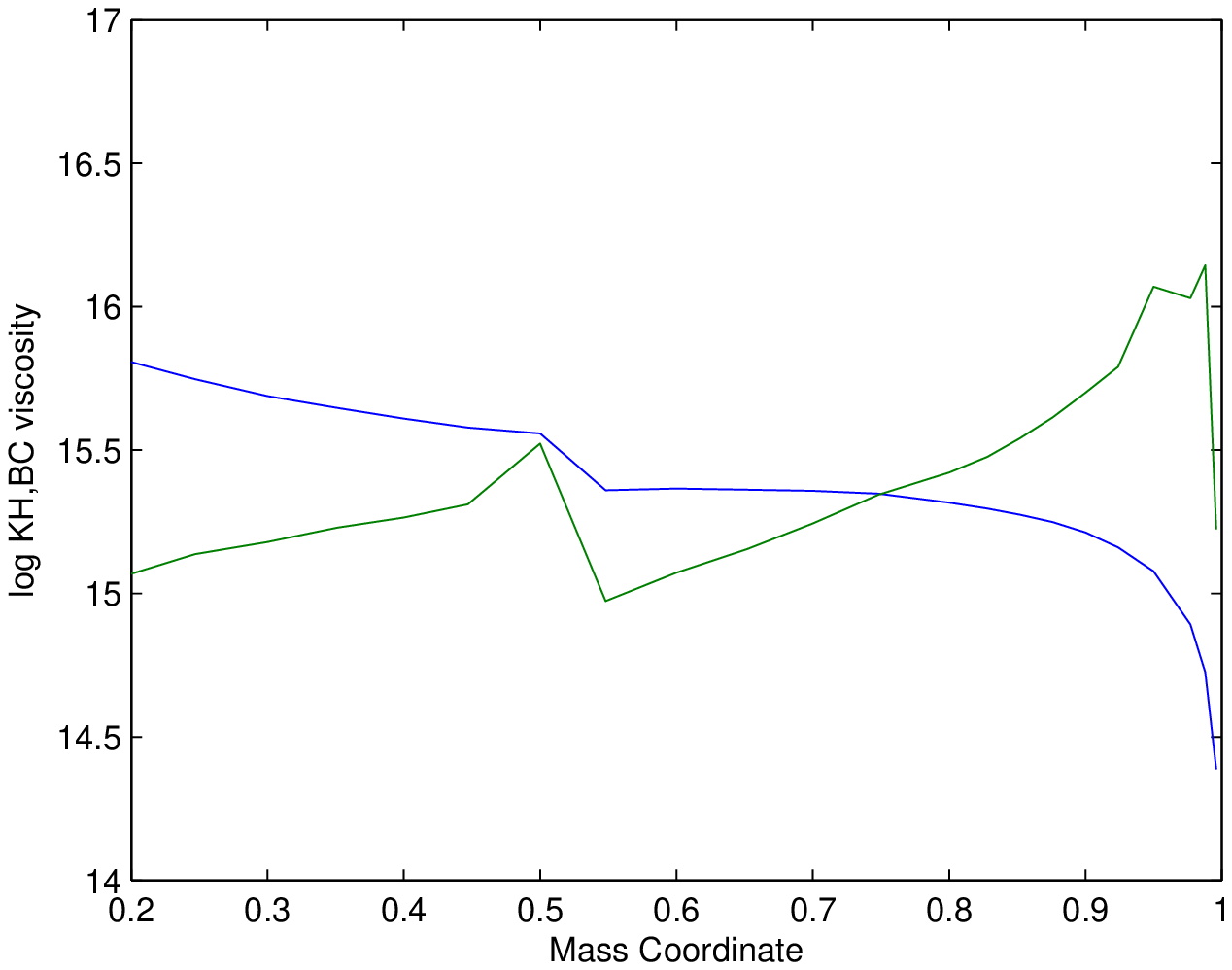}
             {scale=0.5}{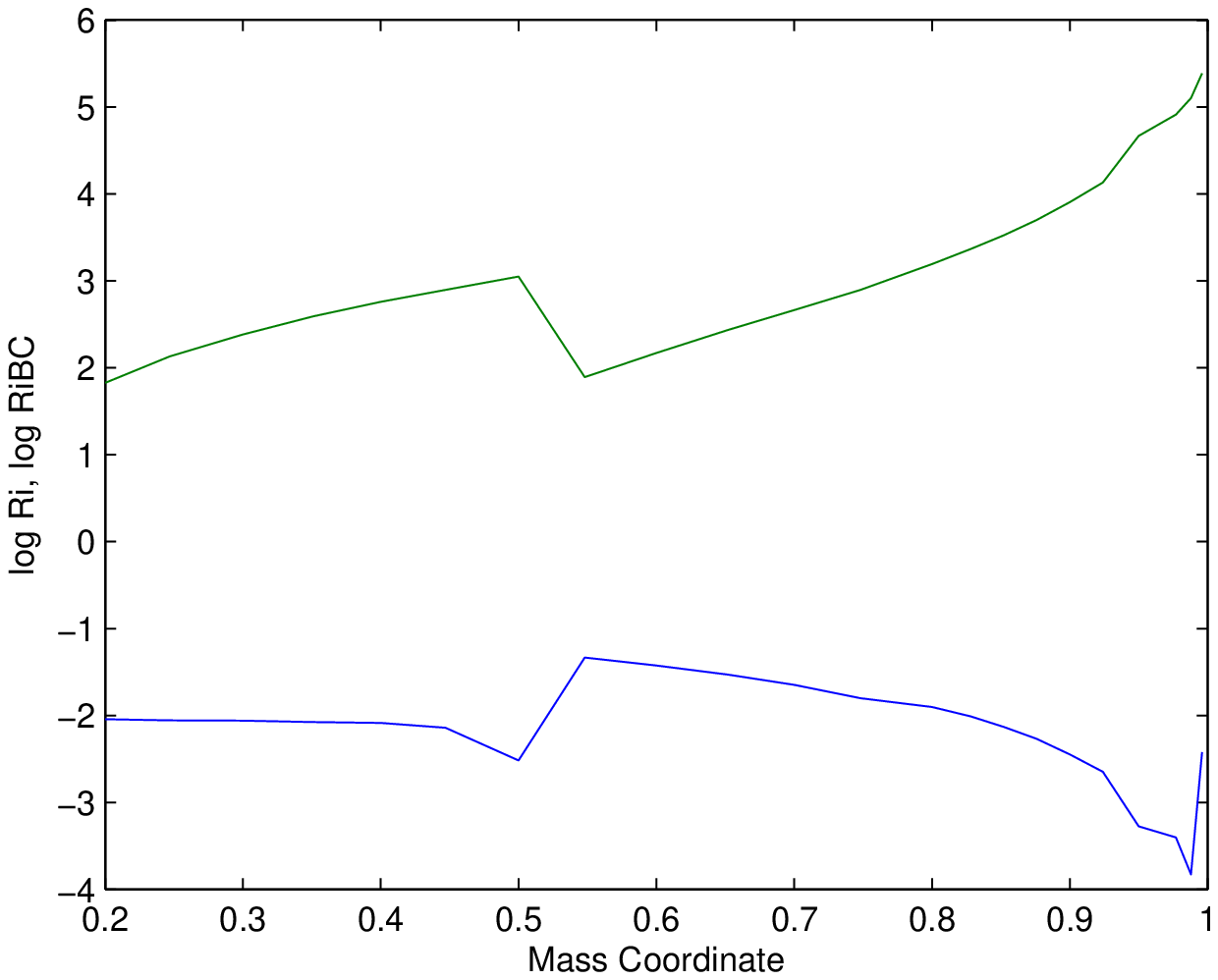}
{\it Profiles of viscosity and Richardson number for the differential
rotation case at the final time-step, calculated self-consistently
during the run. Left Panel: logarithm of the KH and BC viscosities in
cgs units. Color code: KH blue, BC green. Right Panel: logarithm of
${\rm Ri}$ and its critical value ${\rm Ri}_{\rm BC}$ defined in
\S \ref{hydrovisc}. Color code: ${\rm Ri}$ blue, ${\rm Ri}_{\rm BC}$ 
green.}{GWdetail}   

Figure \ref{GWdetail} shows that the KH viscosity is comparable to
or less than the BC viscosity throughout the structure. This suggests
that the solution would not be significantly different if we were to
neglect the BC viscosity and employ only the KH viscosity. We ran
such models and, as expected, found a similar differentially-rotating
solution. See \S\ref{rotyl} for a related discussion.
Note from the right panel of Figure 6 that Ri $\sim 10^{-2}$ in this 
solution, in strong contrast to the value Ri $\sim 10^6$ found in the 
regime of nearly solid body rotation that had negligible KH viscosity, 
and an intermediate value of the BC viscosity.
           
\section{Regime of Internal Rotation for Zahn Viscosity}
\label{rotzahn}

The secular, low, Zahn viscosity (Equation (\ref{Zvisc})) produces 
a regime of differential rotation \citep{sainom}. We show the evolution 
of the $\Omega$-profile as accretion proceeds for the case of Zahn 
viscosity in Figure \ref{ZahnOmega}. The left panel displays a 3-D 
surface plot and the right panel shows stacked profiles in a 2-D plot,
as before. We stress that the viscous timescale $t_{visc}$ in this 
low-viscosity case is generally comparable to the accretion timescale 
$t_{ac}$, a point we discuss in detail in \S \ref{discussregime}. 
The time axis in Figure \ref{ZahnOmega} is marked in units of the 
viscous timescale.

\insertdoblfig{scale=0.5}{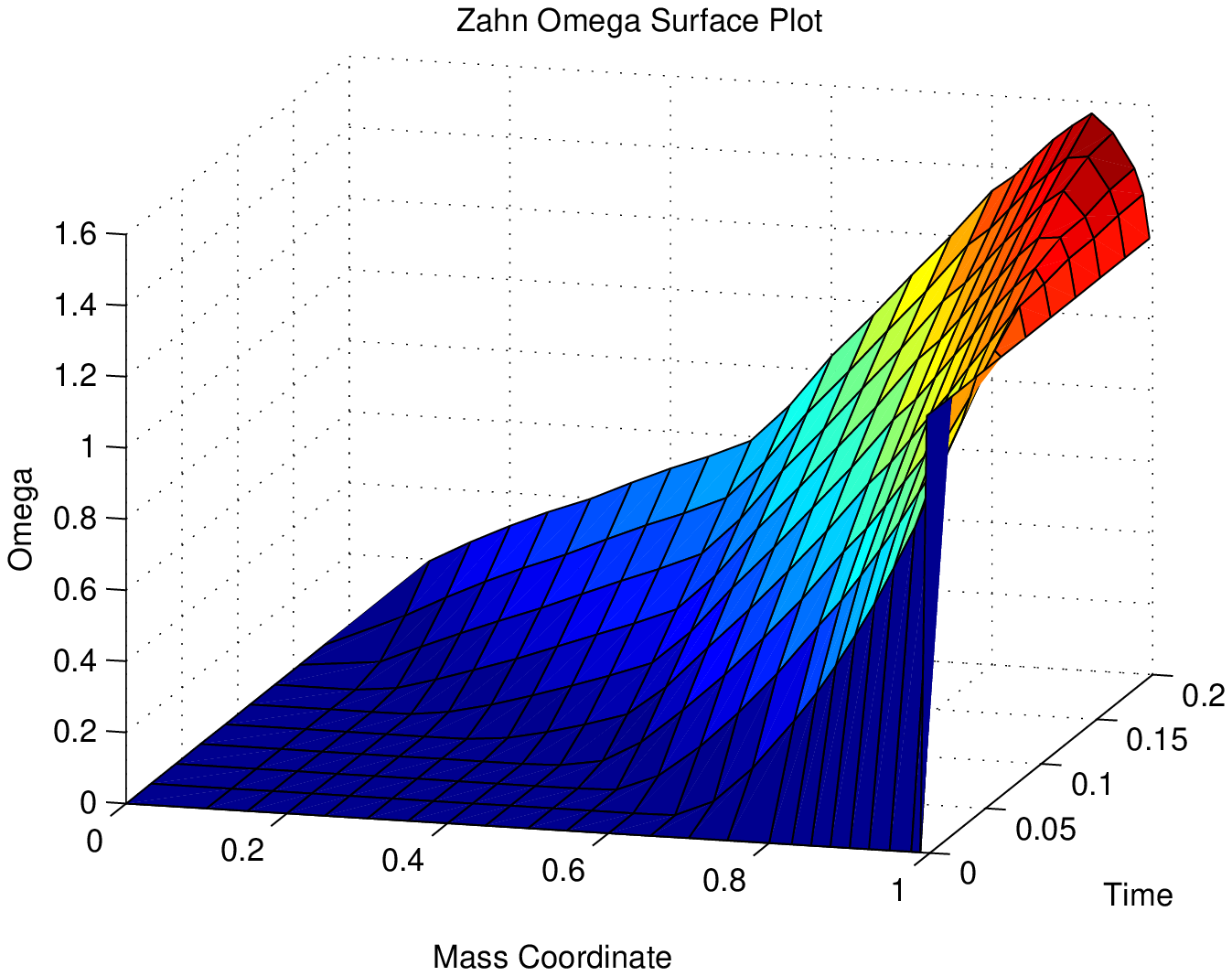}
             {scale=0.5}{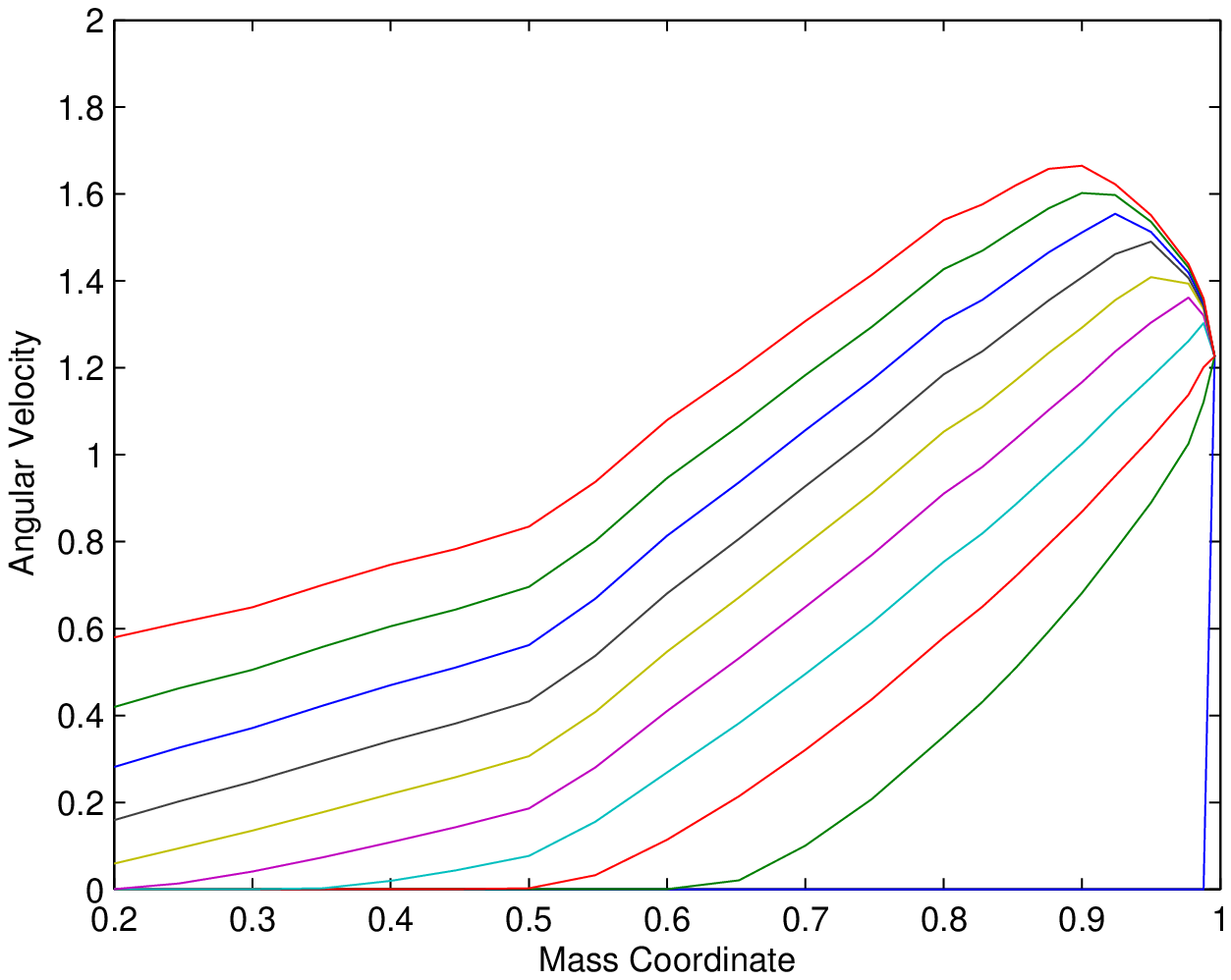}
{\it Evolution of the angular velocity profile
(in units of s$^{-1}$) with Zahn viscosity. 
The time axis is marked in units of the viscous timescale, which
is generally comparable to the accretion timescale for the Zahn
viscosity. Left Panel: 3-D surface plot. Right Panel: Stacked profiles 
in a 2-D plot, wherein the profiles at 10 equal time-steps are color 
coded, with time going up vertically.}{ZahnOmega}  

The evolved Zahn profile has a characteristic maximum in  
$\Omega$ in the outer parts of the WD, with $\Omega$ decreasing
with increasing $r$ or $M_r$ in those parts of the star beyond this 
maximum. The left panel of Figure \ref{Zahndetail} shows the profile 
of the Zahn viscosity \viscZ\ (in cgs units) and the right panel 
that of the Richardson number, Ri, at the final time step, 
calculated self-consistently during the computation. 
Note that Ri is in the low-Ri regime as defined in this work.
The viscosity in this case is also dependent on the shear, through 
the Richardson number, as in the high-viscosity case described earlier.
Figure \ref{Zahndetail} shows that the Zahn viscosity falls in a 
typical range of $10^{2}$ to $10^{4}$ cm$^2$s$^{-1}$, defined here
as a regime of {\sl low viscosity}.

We re-emphasize here that $t_{visc}$ is generally of the same
order as $t_{ac}$ for the low Zahn viscosity. Accordingly, the 
picture presented in \S \ref{diff} for the situation $t_{visc} 
\ll t_{ac}$, wherein the rotation profile can be thought of as 
attaining an ``asymptotic'' shape quickly on the viscous timescale
and that shape then evolving much more slowly on the accretion 
timescale, does not apply here. Instead, the whole profile evolves 
slowly on a common timescale $\sim t_{visc} \sim t_{ac}$. We return 
to this and related matters in \S \ref{discussregime}.  

\insertdoblfig{scale=0.5}{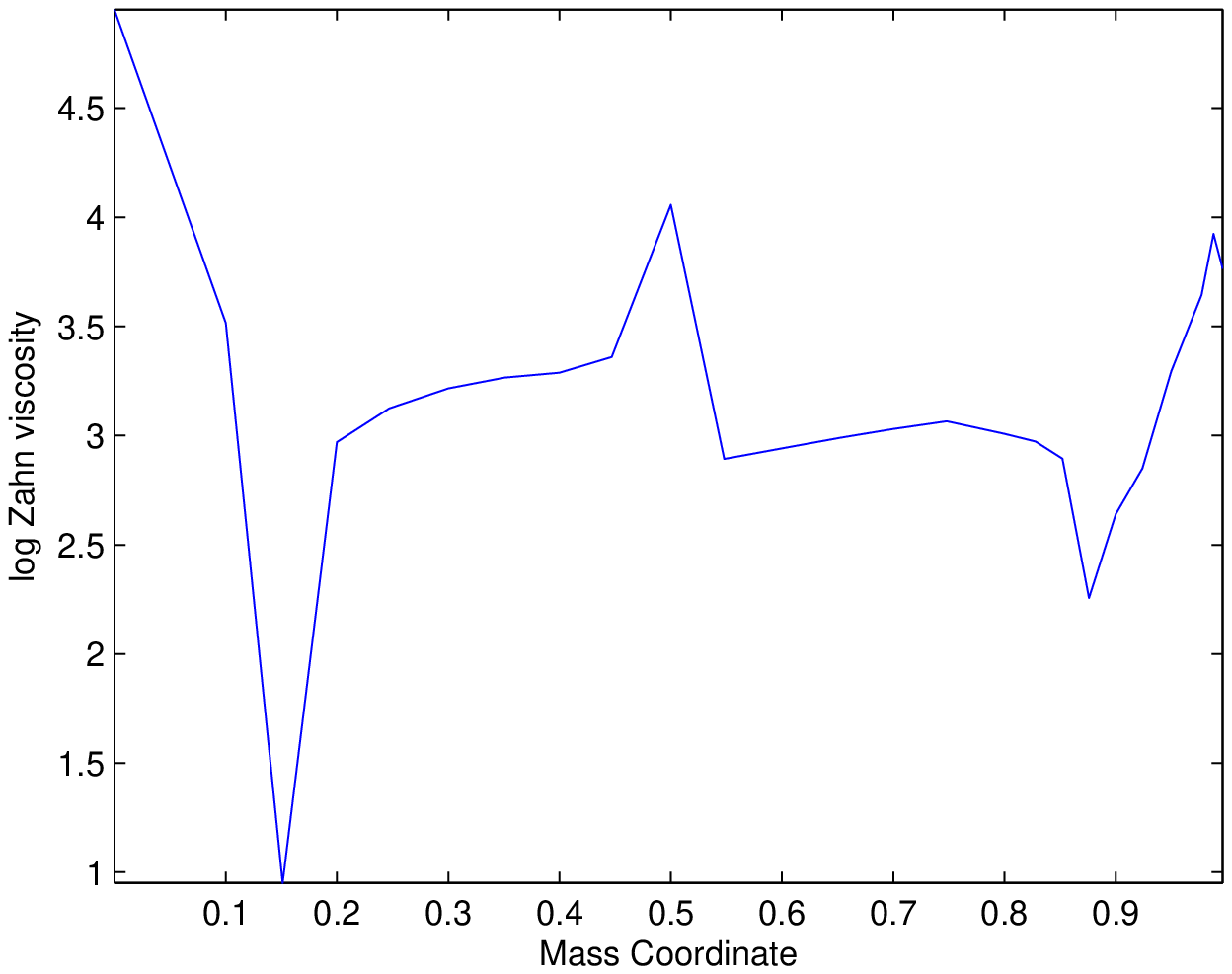}
             {scale=0.5}{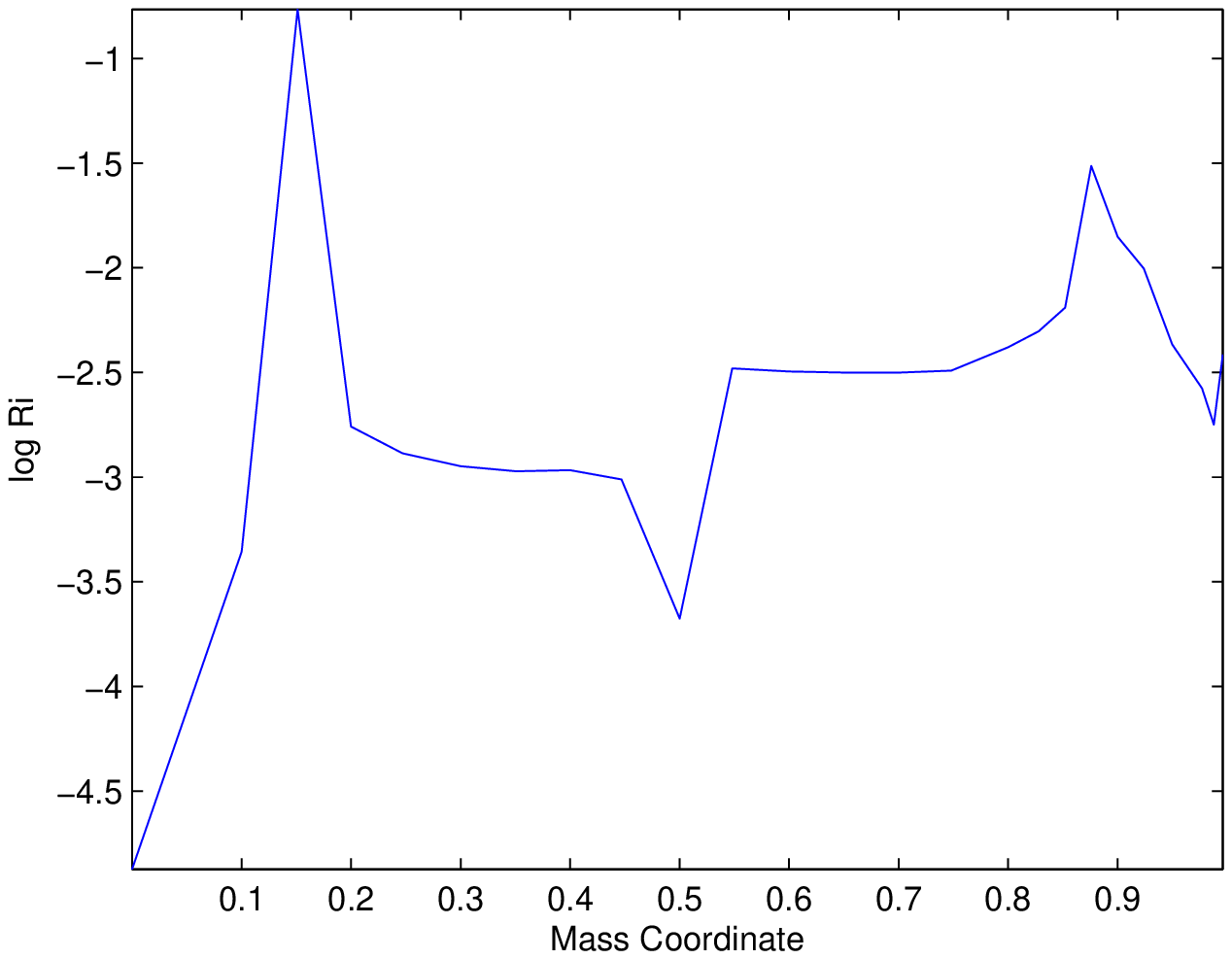}
{\it Profiles of Zahn viscosity and Richardson number at the final 
time-step, calculated self-consistently during the run. Left Panel: 
logarithm of Zahn viscosity in cgs units. Right Panel: logarithm of 
the Richardson number ${\rm Ri}$.}{Zahndetail}    

\section{The Yoon-Langer Regime}
\label{rotyl}

Yoon and Langer (2004, 2005, henceforth YL) studied internal rotation 
profiles of accreting white dwarfs in a pioneering study contemporaneous 
with that of Saio and Nomoto (2004). Among the high, hydrodynamic 
viscosities discussed in \S \ref{hydrovisc}, these authors included
KH viscosity, but not BC viscosity, arguing that it was unlikely 
to be operational inside WDs (see \S\ref{BCI} and \S\ref{discuss}). 
YL also included the low, secular Zahn viscosity described in \S\ref{secvisc} 
and \S\ref{rotzahn}. These authors also included mechanisms we characterized in 
\S\ref{elltransport} as transporters of angular momentum in the \horihat-direction 
or horizontal direction, \viz, ES circulation and GSF instability. 
A typical rotation profile found by YL is shown in Figure \ref{YLOmega}.    

\insertfig{scale=0.5}{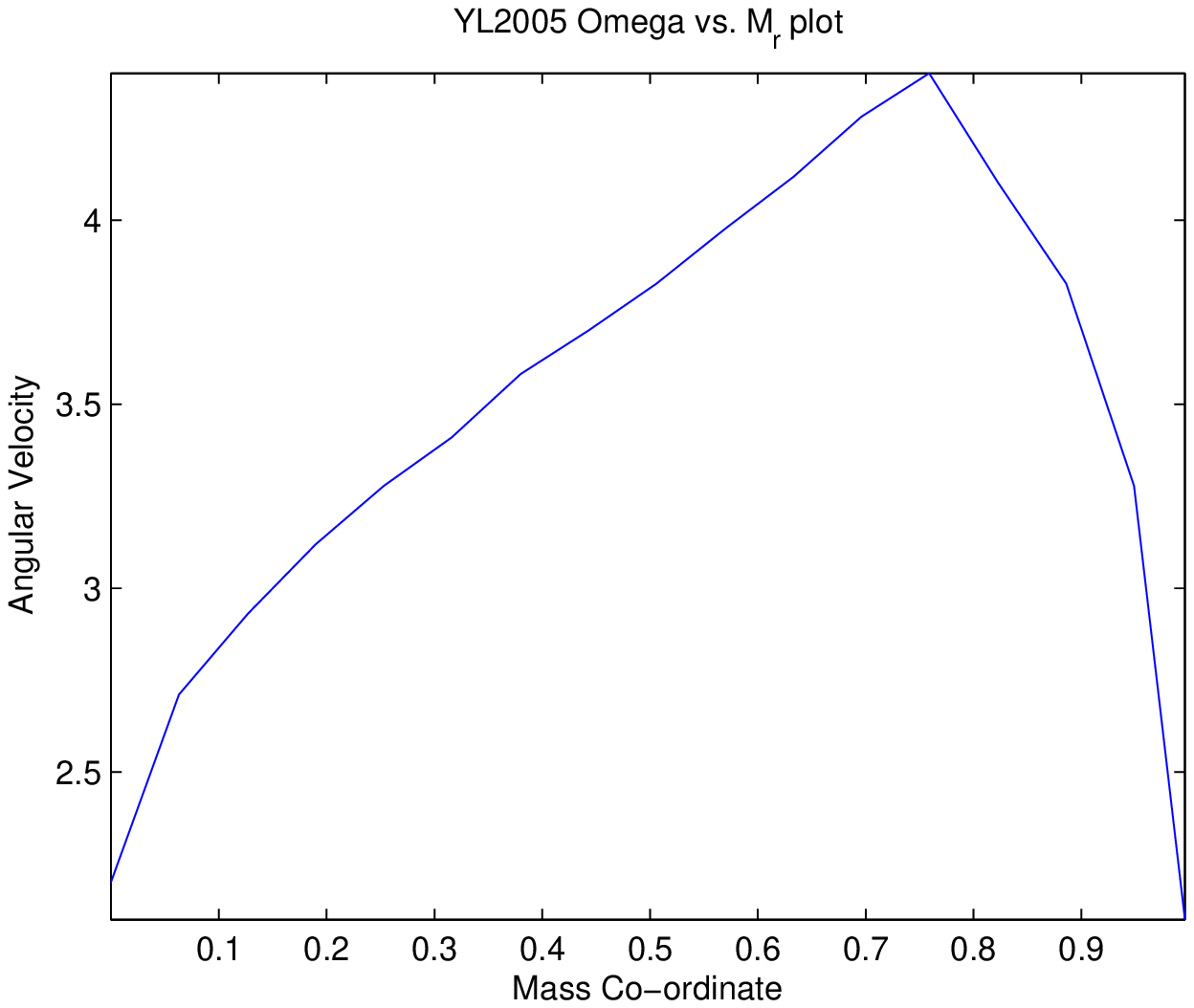}{\it Angular velocity
profile of a typical YL model, showing differential rotation with
a maximum in $\Omega$ (in units of s$^{-1}$)
 in the outer parts of the WD. From Yoon and
Langer (2005).}{YLOmega} 

The YL profile has a characteristic maximum in $\Omega$ in the 
outer parts of the WD, with $\Omega$ decreasing with increasing $r$ 
or $M_r$ in those parts of the star beyond this maximum. At radii
smaller than that corresponding to the maximum in $\Omega$, the
angular momentum transport is dominated by the KH viscosity, 
with the Richardson number Ri maintained very close to its 
critical value of 1/4 (see \S \ref{KHI}) throughout almost all
this inner region. 
The effective viscosity in the inner regions of this solution
is $\nu_{KH} \sim 10^8$ and hence in the range of intermediate
viscosity as defined here.
At larger radii, Ri exceeds its critical value, KHI is
shut off, and the radial angular momentum transport is dominated 
basically by the low Zahn viscosity, with ES and GSF processes
(which generate low viscosities) being also operational. 

The YL regime is thus a combination of two regions, one with  
intermediate viscosity and the other with low viscosity, as opposed to all 
other regimes described in earlier sections of this paper, which 
were governed either entirely by high viscosity, entirely by intermediate
viscosity, or entirely by low viscosity. This leads to interesting consequences, 
which we discuss in \S \ref{discussregime}.  

\section{Multiplicity of Regimes: General Discussion}
\label{discussregime}

We now give a general overview of the multiple rotation regimes 
for accreting WDs found so far, as described above, clarifying the 
current understanding of the origins of these regimes. Figure \ref{Ri_nu}
gives a schematic summary of the regimes we summarize below.

\insertfig{scale=0.5}{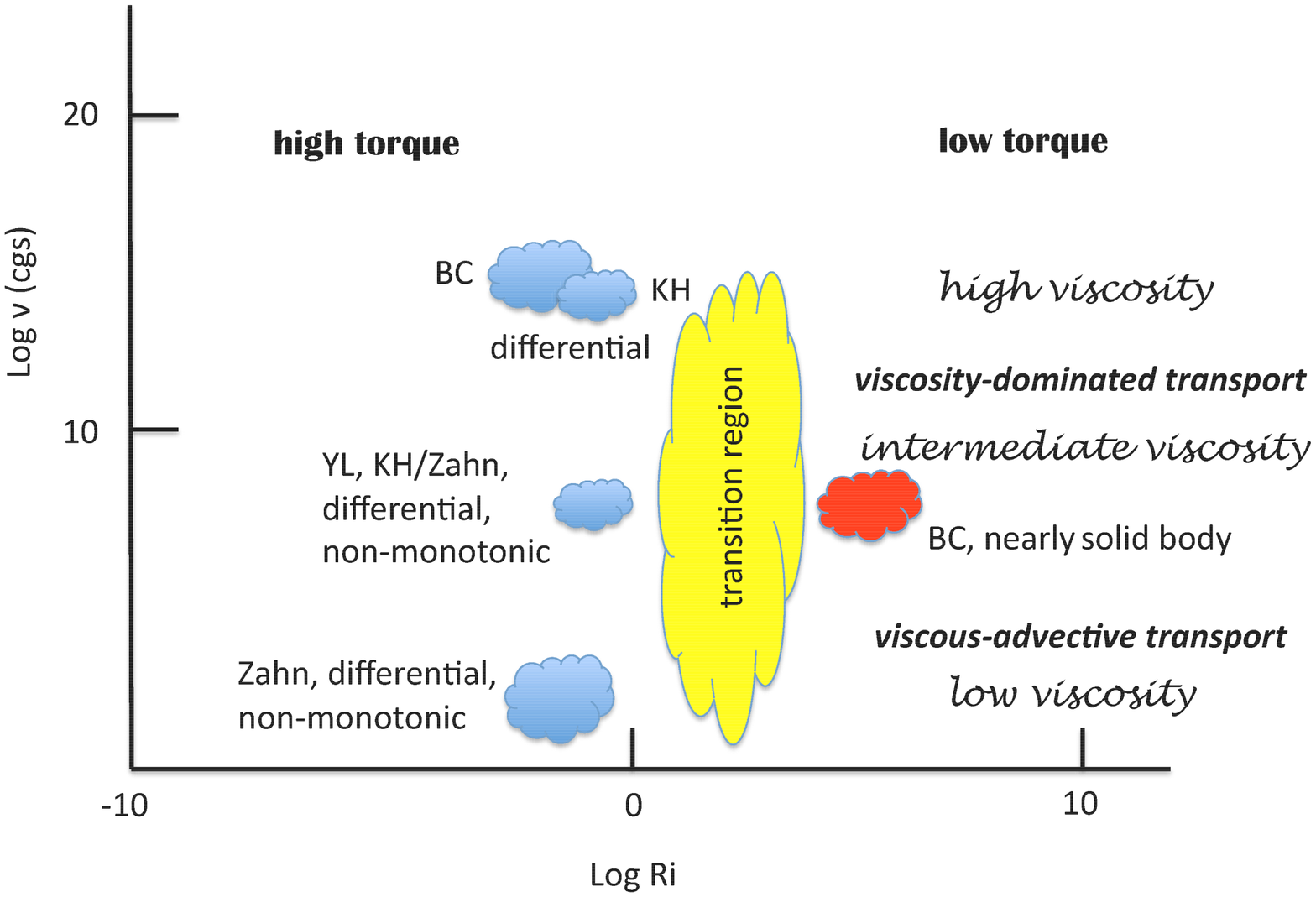}{\it Schematic representation of the
various regimes of rotational state presented in the plane of the log 
of effective viscosity versus the log of Ri.}
{Ri_nu} 

We note first that three timescales are of crucial importance in this 
problem. The first, and by far the shortest, timescale is the viscous timescale
for angular momentum penetration into the whole WD, $t_{KHBC}$, due 
to the large hydrodynamic viscosity, $\nu_{KHBC}$, generated by the 
combined action of KHI and BCI, as detailed in \S \ref{hydrovisc}. 
Since the scale of this viscosity is very large, $\nu_{KHBC}\sim 
10^{15}-10^{16}$ cm$^2$ s$^{-1}$, as seen from Figures \ref{GWdetail}
and \ref{Ri_nu}, the viscous timescale, given by
\begin{equation}
t_{KHBC} \sim \RWD^2/\nu_{KHBC} \sim (1-10)R_8^2 \ {\rm s}\ ,
\label{tGW}
\end{equation}  
is very short. Here, $R_8$ is \RWD\ in units of $10^8$ cm.

The second, much longer timescale, is the viscous timescale for angular 
momentum penetration into the whole WD, $t_Z$, due to the low, secular, 
Zahn viscosity, $\nu_Z$ described in \S \ref{secvisc}. Since the scale of this 
viscosity is very low, $\nu_Z\sim 10^1 - 10^4$ cm$^2$ s$^{-1}$, as seen from 
Figures \ref{Zahndetail} and \ref{Ri_nu}, the viscous timescale, given by
\begin{equation}
t_Z \sim \RWD^2/\nu_Z \sim (10^{12}-10^{15})R_8^2 \ {\rm s}\ ,
\label{tZ}
\end{equation}  
is very long. 

The third timescale is that on which the advection effect, as 
detailed in \S \ref{angmomtransport}, transports angular momentum
into the whole WD. The timescale for this process is simply the 
accretion timescale,
 \begin{equation} 
t_{ac} \equiv \frac{\MWD}{\Mdot_{WD}}  \sim 10^{12} - 10^{14} 
\ {\rm s} \sim 10^5 - 10^7 \ {\rm yr} 
\label{tac}
\end{equation}
in this problem. In the subsequent discussion, we take a 
representative value of $10^{13}$ s where required for explicit 
calculations. 

We note that there is a huge separation between $t_{KHBC}$ and $t_{ac}$,
while there is a large overlap between $t_Z$ and $t_{ac}$. A
straightforward characterization of the rotation regimes that emerge 
from the interplay of these timescales can be given as follows.

Consider first the situation where the hydrodynamic viscosity is
relevant. Then the viscous timescale, $t_{KHBC}$, determines the
fast, viscous evolution, leading to an asymptotic profile that
can be either nearly-uniform or strongly differential rotation 
depending on the operational regime of Ri, as detailed in 
\S \ref{rothydro}. This profile subsequently evolves very slowly 
on the accretion timescale $t_{ac}$. We can call this situation the 
\emph{viscosity--dominated} transport of angular momentum
(Figure \ref{Ri_nu}) subject to the caveats below.  

The two regimes of rotation found in \S\ref{uniform} and \S\ref{diff}, 
\ie, nearly-uniform and strongly differential rotation, arise from 
the fact that there are two regimes of BC viscosity, as shown by 
Equation (\ref{BCvisc}) and illustrated in Figure \ref{Ri_nu}, where the 
scalings of \viscBC\ and the viscous stress, $\tau$, with the shear, 
$\sigma$, are quite different. The former regime is that of 
${\rm Ri} < {\rm Ri}_{\rm BC}$, which corresponds to 
strong differential rotation, and the latter that of ${\rm Ri} \gg 
{\rm Ri}_{\rm BC}$, which corresponds to nearly-uniform rotation.
Note that the viscous stress-shear relation is 
non-linear in both regimes, but the degree of non-linearity is 
different in the two regimes, with $\tau$ scaling as $\sigma^2$ in
the former regime, but as $\sigma^4$ in the latter. We emphasize
that it is this difference in the degree of non-linearity that
causes the two different regimes of rotation. 
A similar argument holds for $\nu_{KH}$ in the regime
of small Ri, in accord with Equation (\ref{KHvisc1}). For large
values of Ri, $\nu_{KH}$ vanishes, of course, as the KHI shuts off.  

A useful way of classifying these two regimes is obtained 
from considerations of angular momentum and torque introduced in 
\S\ref{angmomtransport} through Equations (\ref{transport1}) and 
(\ref{totrate}), when applied to the idea of the two-phase evolution 
of the $\Omega$--profile, first on the viscous then on the accretion 
timescale, for high hydrodynamic viscosities (see the beginning of 
\S\ref{diff}). The profile first evolves on the short viscous timescale 
to an asymptotic, quasi-steady state that is described by the limit 
of Equation (\ref{transport1}) where the left-hand side is approximately zero, 
which implies that $\partial\Sigma/\partial r \approx 0$, \ie, a 
spatially-constant $\Sigma$. This asymptotic viscous state is
thus one in which the total torque $\Sigma$ is roughly a constant 
inside the white dwarf, which we can label $\Sigma_0$. As accretion proceeds 
on a much longer timescale, stellar conditions evolve slowly on this long 
accretion timescale, and $\Sigma_0$ also evolves on the accretion
timescale, corresponding to the slow evolution of the asymptotic profile 
on the accretion timescale. 

The two regimes can now be described in terms of the two
constant-torque asymptotic states in a straightforward way. Note first that 
the total torque $\Sigma$ is given by Equation (\ref{totrate}), the first term on the 
right-hand side being the viscous part and the second, the advective
part. In the high-viscosity regime, the first term is normally dominant. 
Consider first the strong differential-rotation regime, where the shear is 
large, Ri is small, the viscosity is high, and the viscous term is thus
completely dominant. Here the constant torque is given approximately by 
$ \Sigma_0 \approx 4\pi\rho r^4\nu(\partial\Omega/\partial r)$, a large 
constant with a finite shear. Thus, there is strong angular momentum transport 
in this asymptotic state, dominated by viscous transport. Consider next the 
nearly-uniform rotation regime, where the shear is small, Ri is large,
and the viscosity takes an intermediate value as illustrated in Figure \ref{Ri_nu}.
Here, both viscous and advective components are small, and so, therefore, 
is the constant torque, $\Sigma_0$.  In this small-torque regime, the 
angular momentum transport is very small, but non-vanishing. 

A constant-torque solution corresponding to the asymptotic viscous profile 
is thus a generic feature of the high-viscosity regimes. The case where the 
constant, $\Sigma_0$, is very small is the nearly-uniform rotation regime, 
which does not appear to be otherwise special. We emphasize that a 
{\it strictly} constant-$\Omega$ solution is not possible, as there would be 
no viscous transport at all in that case: the shear would then be exactly 
zero, and so would be both the KH and the BC viscosities, as Equations 
(\ref{KHvisc1}) and (\ref{BCvisc}) show.  

A final remark about the hydrodynamic viscosities used in this work 
is that there may be caveats about their prescriptions that need to 
be studied further. We briefly consider here those relating to the 
BC viscosity. The widely-used Fujimoto prescription
\citep{fuji93,sainom,piro}, as given by Equation (\ref{BCvisc}), is 
obtained in the limit Ri $\gg$ 1 \citep{fuji93}. For the regime of 
rotation profile described by Figure \ref{GWdetail}, however, this condition is 
not satisfied. In order to check the self-consistency of this regime, we 
have therefore gone back to the original Fujimoto (1988) prescription
(see Equation (26) of that paper),
which is more complicated but valid for general values of Ri, and 
calculated values of \viscBC\ corresponding to the range of Ri-values 
shown in Figure \ref{GWdetail}. We have then compared these \viscBC -values 
with those in Figure \ref{GWdetail}, obtained with the simple prescription
of Equation (\ref{BCvisc}). We have found that the full prescription gives 
\viscBC -values that are smaller than those obtained from the simple
prescription by factors that range between 2 and 5 over the entire
range of Ri-values given in Figure \ref{GWdetail}. We recall now from
\S\ref{diff} that the rotation profile in this regime is determined by
the sum of KH and BC viscosities, and that the two are generally 
comparable in value, with \viscKH\ having a higher value over most of
the star, as Figure \ref{GWdetail} shows. Therefore, the rotation profile 
obtained with the more complicated BC-prescription will be qualitatively
very similar to that obtained with the simpler prescription, although
it will have a higher dominance of KH viscosity. Thus this regime is 
roughly self-consistent. We do not adopt the more complicated Fujimoto 
prescription of BC viscosity in this work because of other, more 
general, concerns about the BC viscosity (\S\ref{discuss}).

Consider now the situation when the secular Zahn viscosity is relevant, 
Figure \ref{Ri_nu}. Since $t_{ac} \sim t_Z$ over most of the phase space, 
viscosity and advection generally play comparable roles in angular 
momentum transport in this case. The result is a regime of
differential rotation where the penetration of angular momentum into 
the interior of the WD is due to both viscosity and advection, the 
two terms being roughly comparable in magnitude over most of the 
WD interior. We call this situation the {\sl viscous--advective} 
transport of angular momentum (Figure \ref{Ri_nu}). Similar or 
related situations have been studied in other areas of physics and 
are sometimes called {\sl advection--diffusion}. 

Consider, finally, the YL regime that requires a detailed  discussion 
because of its special, interesting nature, as indicated in \S \ref{rotyl}. 
We show first that the hierarchy of timescales established at the 
beginning of this section also holds for the YL regime. This is so 
because the viscous timescale for angular momentum penetration into the WD, $t_{KH}^{YL}$, due 
to the large hydrodynamic viscosity $\nu_{KH}^{YL}$ generated 
by KHI in the YL regime pertains in the inner regions of the WD. 
This regime of intermediate viscosity and short viscous time occurs 
at radii smaller than that at which $\Omega$ has a maximum (this 
region covers roughly the inner $\sim 80 \%$ of the WD mass). The
scale of $\nu_{KH}^{YL} \sim 10^8-10^9$ cm$^2$ s$^{-1}$ can be
seen in Figure 8 of Yoon and Langer (2004). This is quite large,
although smaller than the high-viscosity scale associated with the 
regime of low Ri (Figure \ref{Ri_nu}), and hence in the range of
intermediate viscosity as defined here. The reduction in viscosity 
is due to the fact that the Richardson number is very close to its critical value 1/4 
(see Equations and (\ref{KHvisc1}) and (\ref{KHvisc2})) 
over this entire region of the YL regime. This value of the KH
viscosity leads to a viscous timescale given by
\begin{equation}
t_{KH}^{YL} \sim \RWD^2/\nu_{KH}^{YL}
\sim (10^7-10^8)R_8^2 \ {\rm s}\ ,
\label{tYL}
\end{equation} 
which can be compared with Zahn timescale $t_Z$ and the 
accretion timescale $t_{ac}$, which are still given by 
Equations (\ref{tZ}) and (\ref{tac}) respectively. 

The nature of the YL regime can 
now be clarified in a straightforward way. 
As emphasized by these authors, although the inner region of
their regime does have a large, hydrodynamic viscosity 
and so a rapid viscous transport, the low Zahn viscosity 
in their outer region with its slow transport (with a 
comparable contribution from advection) represents a severe 
``bottleneck'' in the angular momentum transport into the WD 
interior \citep{yl04}. It could be argued, therefore, that
overall penetration of the accreted angular momentum into 
the WD can only happen on the long timescale $\sim t_{ac} 
\sim t_Z$. One could then argue further that the Zahn and YL 
regimes are rather similar in this sense, and cite the general 
similarity between the shapes of the Zahn and YL 
$\Omega$--profiles (even to the extent that the maximum in 
$\Omega$ comes in the range  $\sim 80-90 \%$ of the WD mass 
in both cases) in support of this point of view. Strictly 
speaking, however, this is not correct because there is provision 
for fast viscous transport in the YL case in the deep interior of 
the WD, once the angular momentum reaches there. Indeed, in
analogy with the discussion given earlier, the question of a 
possible ``asymptotic profile'' in this interior YL region 
merits detailed study, but is outside the scope of this paper.

We thus see that, among the multiple regimes of rotation 
described above, many possible regimes of differential rotation 
exist, in addition to that of uniform rotation. Indeed, it 
appears that the \emph{generic} regime is one of differential 
rotation and the uniform rotation regime may be a special case.

\section{Discussion and Conclusions}
\label{discuss}
   
We have attempted to clarify the nature of the different regimes of 
internal rotation that can exist inside accreting WDs. We consider 
WDs without strong, permanent magnetic fields and hence consider angular-momentum 
transport processes described by viscosities generated by hydrodynamic mechanisms 
such as Kelvin--Helmholtz and baroclinic instabilities, and secular mechanisms such 
as the Zahn instability. We formulated these viscosities with the aid of 
prescriptions given in the literature.

We elucidate that the BC viscosity as adopted here implies two asymptotic regimes 
depending on the Richardson number, Ri. The BC viscosity, \viscBC, scales 
as ${\rm Ri}^{-1/2} \propto \sigma$ in the regime of small Ri. This yields a 
regime of high viscosity that nevertheless corresponds to differential rotation,
a solution presented here for the first time. The BC viscosity scales as 
${\rm Ri}^{-3/2} \propto \sigma^3$ in the regime of large Ri, yielding an
intermediate level of viscosity and solid-body rotation.

We thus classify the collection of rotation regimes explored here in terms of 
the Richardson number, Ri, in the following way. At small values of Ri $\lta$ 0.1, 
we have both the low-viscosity Zahn regime and the new, high-viscosity regime 
found in this work. Near the critical value of Ri $\approx$ 1/4, we have the 
inner, intermediate-viscosity region of the YL regime, its outer, low-viscosity region 
being at a higher value of Ri. Employment of KH viscosity alone yields 
differential rotation. Finally, at large values of Ri $\gg$ 1, we have the 
intermediate-viscosity nearly-uniform rotation regime. 

The two regimes of the BC viscosity correspond to a viscosity that scales 
with the shear as $\sigma$ in the small Ri regime and as $\sigma^3$ in the large 
Ri regime (\S\ref{BCI}). This implies, in turn, that the viscous stress, 
$\tau = \nu \sigma$, scales with the shear as $\sigma^2$ in the small Ri regime, but 
as $\sigma^4$ in the large Ri regime. The form of nonlinearity is a power-law with
an index $n>1$, $n$ being 2 in the small Ri regime, and 4 in the large Ri
regime. It is interesting to consider what the situation would be in a more 
complete theory of BC viscosity. Instead of the knowledge of \viscBC\ only 
in the two limits ${\rm Ri} < {\rm Ri}_{\rm BC}$ and ${\rm Ri} \gg {\rm Ri}_{\rm BC}$,   
we would then have a complete prescription over the whole range of Ri, 
which would reduce to the above limits appropriately. The non-linearity 
in the viscous stress will change from a power-law with $n=2$ to one with 
$n=4$ as Ri increases, and the rotation regime will change from a strongly 
differential to nearly-uniform. Accordingly, the two limiting regimes will 
have a \emph{transition region} between them, as illustrated schematically
in Figure \ref{Ri_nu}. 

An insight that is particularly useful for our problem in the regimes of 
power-law viscosities is available from the literature on the p-Laplacian 
nonlinear diffusion equation \citep{KaVa,lpv,B-V,AkMa}, whose form, 
$\partial u/\partial t = {\bf\nabla}\cdot (\vert{\bf\nabla} u\vert^{p-2}{\bf\nabla} u)$, 
is very similar to that of Equation (\ref{transport2}) in the limit where 
viscous transport dominates over advective transport, as is the 
case for the large hydrodynamic viscosities. This similarity becomes 
clear if the advective term in Equation (\ref{transport2}) is neglected and 
the equation is rewritten back in terms of the angular velocity 
$\Omega$ to read $\partial\Omega/\partial t = (1/\rho r^4)
(\partial/\partial r)(\rho r^4\nu(\partial\Omega/\partial r))$. 
The power-law index $n$ for the viscous stress $\nu \sigma$ is related 
to the index $p$ by $n=p-1$, so that that the two asymptotic regimes 
correspond to $p=3$ and $p=5$, respectively. 

The non-trivial asymptotic solution of the p-Laplacian PDE for a 
\emph{given} value of $p$ can be shown to be unique, but the solutions 
for \emph{different} values of $p$ are quite different \citep{KaVa,lpv}.            
This is consistent with the two viscosity regimes found here.
The implication is that when $p$ changes from one value to another 
through the transition region, the solution will undergo a corresponding 
transition. The specific issue of the behavior of the BC viscosity at 
such intermediate $p$- or $n$-values (or, equivalently, the corresponding 
values of Ri) are beyond the scope of this paper, but of definite interest.

There has been discussion in the literature about the applicability 
of BCI in the electron-degenerate matter inside WDs since completely 
degenerate matter is barotropic, pressure is only a function of density
and their gradients cannot be skew to one another (Kippenhahn \& 
M\"olenhoff 1974; Tassoul 1984; Yoon \& Langer 2004). Realistic WDs 
have finite temperature so the issue of baroclinicity is a quantitative
one. The key point is that buoyancy may become very small in degenerate
matter, with consequences for Ri and other parameters of the problem, in 
which case the standard BCI formulation, including the Fujimoto prescription, 
may not apply. A first-principles reformulation of the problem may be necessary. 

While our immediate goal in this work has been to better
understand the physics that drives the differential rotation regimes
of accreting white dwarfs, our larger, long--term goal is to elucidate 
the progenitor evolution of SN~Ia. Currently viable models of SN~Ia 
involve either accretion onto a WD from a non--degenerate star (the SD 
scenario), from a degenerate companion (the DD scenario), or the violent 
impact of two degenerate stars (Maoz, Mannucci \& Nelemans 2014). All these 
possibilities demand that the WD progenitor of the SN~Ia be rotating. It 
is an interesting task to use the physics of angular momentum transport 
described here to elucidate and guide our understanding of the possible 
rotation state of the progenitors of SN~Ia. A major challenge remains
to determine what rotational states nature accommodates, under what 
circumstances, and to what effect on the subsequent explosion. 

The reverse problem, \viz, using observations of SN~Ia to constrain the 
physics of the progenitor evolution, is equally interesting and perhaps even
more challenging. While acknowledging the definite variety among typical SN~Ia, 
we must recognize that the majority of these events display a remarkable degree of 
homogeneity. The fact that most SN~Ia are consistent with values of ejecta 
mass not remarkably different from \MCh\ suggests that most SN~Ia cannot be 
in a regime of strong differential rotation, neither that of low Ri we have presented
here nor in the regime propounded by Yoon \& Langer (2004). Thus, it is important 
to understand the conditions that prevent regimes of differential rotation from 
occurring in common circumstances. The answer cannot be so simple as saying 
that the viscosity, even magnetic viscosity, is large, because we have shown here 
that there can be differential rotation in a high-viscosity regime. Likewise, some 
SN~Ia fall in the Super-Chandrasekhar regime that would seem to demand 
differential rotation, implying that the nearly solid-body rotation solutions of 
Saio \& Nomoto (2004) and of Piro (2008) do not pertain. If many SN~Ia 
involve sub-Chandra masses, as suggested by Scalzo et al. (2014) and others,
then the state of rotation is currently unconstrained. The nature of internal rotational 
profiles in all these circumstances requires deeper understanding. 

\acknowledgments
We thank Michael Montgomery for providing the basic white dwarf model
and for helpful discussion of white dwarf structure and the 
referee for helpful comments which improved the paper. Some work 
on this paper was done in the hospitable environment of the Aspen Center 
for Physics that is supported by NSF Grant PHY-1066293. P. G. thanks 
the University of Texas at Austin for its warm hospitality during 
part of this work. The authors are also grateful for the Posse East 
establishment where much of the conceptual development of this work 
was done. This work was supported in part by NSF Grant NSF AST-1109801. 

\appendix

\section{The Advection Effect}
\label{advection}

In modeling the inward mass advection rate $\Mdot(r,t)$ in the
interior of the WD, we first note that, due to mass conservation 
inside the WD, we can express the advection rate as
\begin{equation}
\Mdot(r,t) = \dot{M}_r(r,t),
\label{masscons} 
\end{equation}  
in terms of the mass co-ordinate $M_r$ introduced in 
\S \ref{angmomtransport}, and defined by
\begin{equation}
M_r = \int_0^r 4\pi\rho(r,t)r^2dr.
\label{masscoord} 
\end{equation} 
We must then model $\dot{M}_r$ for accreting WDs in
terms of the accretion rate $\Mdot_{WD}$ on the WD, or, equivalently,
model $\dot{M}_r/M_r$ in terms of $\Mdot_{WD}/\MWD$. 

For the simple case of a uniform density profile, it is trivial to
show that 
\begin{equation}
{\dot{M}_r\over M_r} \propto {\Mdot_{WD}\over\MWD}.
\label{unicase} 
\end{equation} 
For approximately self-similar contraction of the WD as its mass
increases, \ie, a contraction during which the density profile 
remains roughly self-similar, Equation (\ref{unicase}) still holds
approximately, with the constant of proportionality replaced by
a slowly--varying function of radius. The value of this function
varies by a factor $\sim 1$ between the center and the surface
of the star, so that Equation (\ref{unicase}) is approximately valid over
the entire extent of the WD. This means that $\dot{M}_r/M_r$
is roughly independent of radius and roughly proportional to 
$\Mdot_{WD}/\MWD$. In other words, the dominant radial variation
in $\dot{M}_r$ is given by that in $M_r$; detailed structure 
effects give only a slow variation.
     
We illustrate this point with polytropic models, used widely in 
studies of WDs. A poytrope with polytropic index $n$ has an 
equation of state $p = K\rho^{1+1/n}$, and a density profile 
$\rho = \rho_c\theta^n$, where $\theta(\xi)$ is the Lane-Emden 
function of index $n$. Here, $\rho_c$ is the central density,
and the scale, $a$, used to define the dimensionless radial 
co-ordinate
\begin{equation} 
\xi\equiv r/a
\label{xidefn} 
\end{equation}
inside the polytrope is given by
\begin{equation}
a \equiv \left[{(n+1)K\rho_c^{\left({1\over n} - 1\right)}\over
4\pi G}\right]^{1/2}.
\label{adefn} 
\end{equation}
The mass of the polytropic WD is given by
\begin{equation}
\MWD \equiv 4\pi \left[{(n+1)K\over
4\pi G}\right]^{3/2}\rho_c^{3-n\over2n}\xi_1^2
\mid\theta ' (\xi_1)\mid,
\label{Mpoly} 
\end{equation}   
and its radius by $\RWD = a\xi_1$. Here, $\xi_1$ is the value 
of $\xi$ at the surface of the polytrope (where the density is
zero), and $\theta ' (\xi_1)$ is the value of the derivative 
of $\theta$ at $\xi_1$. Tables of $\xi_1$ and 
$\mid\theta ' (\xi_1)\mid$ for various values of $n$ are widely
available.

The value of $M_r$, as given by Equation (\ref{masscoord}), can be readily
calculated with the aid of the equations given in the previous 
paragraph, and this value can be related to \MWD\ as given by 
Equation (\ref{Mpoly}) to obtain
\begin{equation}
M_r = {\MWD\phi(\xi,n)\over\xi_1^2\mid\theta ' (\xi_1)\mid}.
\label{MrPoly} 
\end{equation}  
Here, 
\begin{equation}
\phi(\xi,n)\equiv\int_0^{\xi}\theta^n(\xi)\xi^2d\xi
\label{phidefn} 
\end{equation}  
is an appropriate moment of the Lane-Emden function $\theta$.

Equation (\ref{MrPoly}) allows us to calculate the left-hand side of 
Equation (\ref{unicase}) for the polytropic case as
\begin{equation}
{\dot{M}_r\over M_r} = {\Mdot_{WD}\over\MWD} + {\dot{\phi}\over\phi}.
\label{polycase} 
\end{equation} 
The second term on the right-hand side of Equation (\ref{polycase}) can
be evaluated with the aid of Equations (\ref{phidefn}) and 
(\ref{xidefn}) as
\begin{equation}
{\dot{\phi}\over\phi} = - {\theta^n(\xi)\xi^3\over\phi}
\left({\dot{a}\over a}\right).
\label{poly1} 
\end{equation}

The relation between the total mass, \MWD, and the radius scale, $a$,
for a polytrope of index $n$ is well--known to be 
\begin{equation}
\MWD.a^{\left({3-n\over n-1}\right)} = {\rm constant},
\label{poly2} 
\end{equation} 
and can be derived by combining Equations (\ref{adefn}) and (\ref{Mpoly}),
the constant in the above equation coming out to be 
$4\pi\left[(n+1)K\over 4\pi G\right]^{n\over n-1}
\xi_1^2\mid\theta ' (\xi_1)\mid$. Equation (\ref{poly2}) is
closely related to the mass-radius relation for polytropes. 
A logarithmic differentiation of it yields 
\begin{equation}
{\dot{a}\over a} = -\left({n-1\over 3-n}\right){\Mdot_{WD}\over\MWD}.
\label{poly3} 
\end{equation}
Combining Equations (\ref{polycase}), (\ref{poly1}), and (\ref{poly3}), 
we arrive at the final result:
\begin{equation}
{\dot{M}_r\over M_r} = {\Mdot_{WD}\over\MWD}g_n(\xi),
\label{polyfin} 
\end{equation}
where the function $g_n(\xi)$ is defined by  
\begin{equation}
g_n(\xi) \equiv 1 + \left({n-1\over 3-n}\right)
{\theta^n(\xi)\xi^3\over\phi(\xi,n)}. 
\label{gndef} 
\end{equation}

\insertfig{scale=0.5}{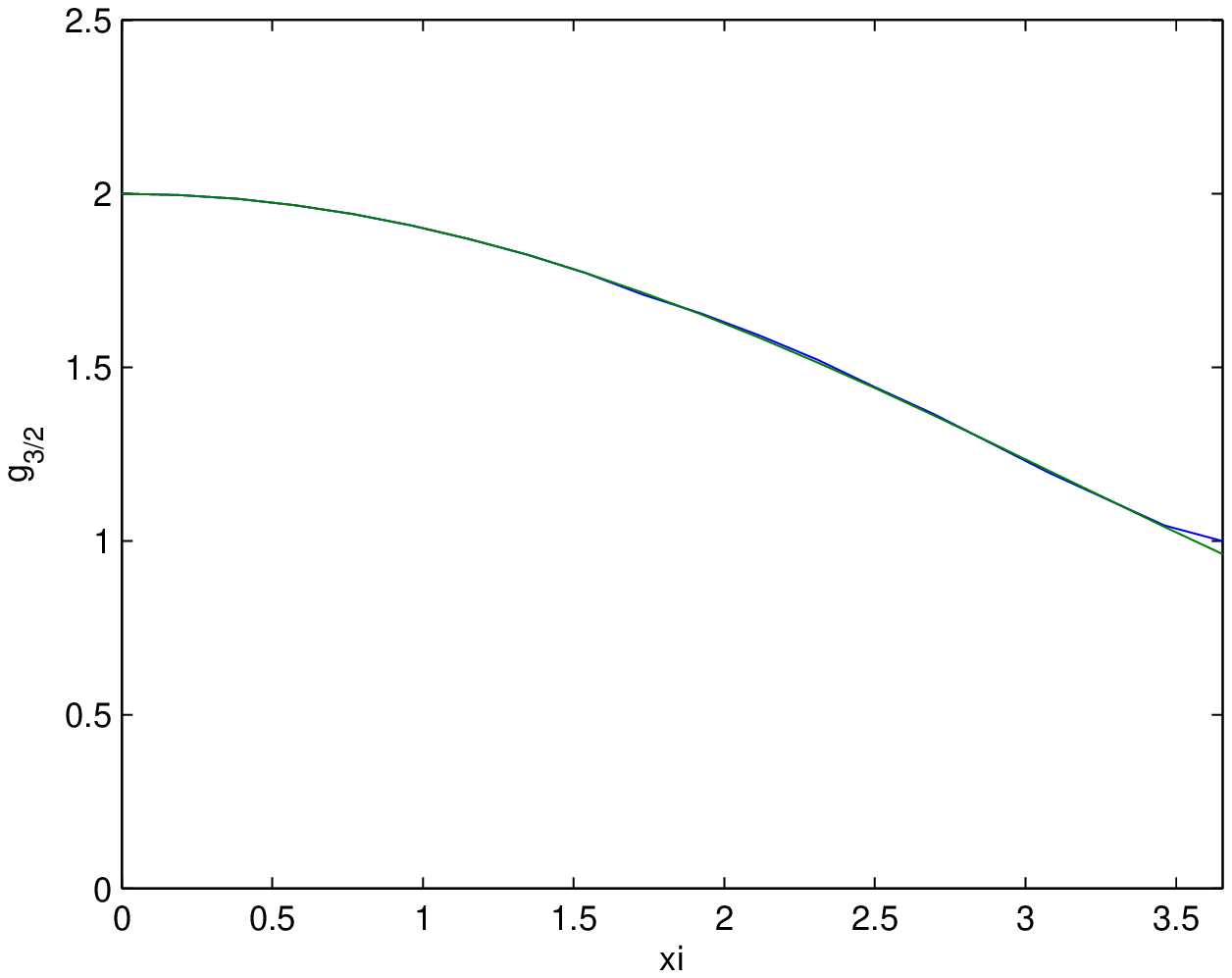}{\it The function $g_{3/2}(\xi)$
for $n=3/2$ polytropes. Note that the surface of the polytrope
(where the density falls to zero) corresponds to $\xi = \xi_1
\approx 3.654$ for this value of the polytropic index. Blue line:
Exact calculation. Green line: Taylor approximation.}{advec}

The function $g_{3/2}(\xi)$ for the polytropic index $n=3/2$, often
used in model WD calculations, is shown in Figure \ref{advec}. It is
a slowly--varying functon, increasing from unity at the surface of
the polytrope to a value of 2 at the center. For comparison, we 
have used a Taylor expansion of the Lane-Emden function of 
polytropic index $n$, which is given by      
\begin{equation}
\theta_n(\xi) = 1 - {1\over 6}\xi^2 + {n\over 120}\xi^4 
-{n(8n-5)\over 15120}\xi^6 + \dots ,
\label{taylor1} 
\end{equation}
to derive a series representation of $g_n$, the result being
\begin{equation}
g_n(\xi) = 1 + {3(n-1)\over 3-n}\left[1 - {n\over 15}\xi^2 +
{n(19n-25)\over 3150}\xi^4 - \dots \right] ,
\label{taylor2} 
\end{equation} 
which reduces in the special case $n=3/2$ to
\begin{equation}
g_{3/2}(\xi) = 2 - {\xi^2\over 10} + {\xi^4\over 600} - \dots .
\label{taylor3} 
\end{equation}
This Taylor approximation to $g_{3/2}(\xi)$ is also shown in 
Fig.\ref{advec}, keeping only the terms given in 
Equation (\ref{taylor3}), and showing thereby how close this 
approximation is to the exact value.

Similar results are obtained for other polytropic indices,
leading us to use the approximation for the advection rate
given by Equation (\ref{advrate}) in \S \ref{angmomtransport}.

\section{YL Boundary Condition: A Simple Prescription}
\label{YLbcsimple}

As explained in \S \ref{diff}, we envisage angular momentum being 
added to a non-rotating WD, so that the specific angular momentum 
at its surface, $\ell_{surf}$, increases according to the 
prescription
\begin{equation}
{\partial\ell_{surf}\over\partial t} = {\ell_K(\RWD)\over t_{ac}},
\label{YLs1}
\end{equation}
where
\begin{equation}
\ell_K(\RWD) = \sqrt{G\MWD\RWD}
\label{ellKsurf}
\end{equation}
is the Keplerian value of $\ell$ at the surface of the WD of mass 
\MWD\ and radius \RWD. We emphasize that both \MWD\ and \RWD\
here are functions of time as accretion goes on, which is of 
crucial importance in the relations given below. 

We use the mass-radius relation 
\begin{equation}
\RWD \propto \MWD^{-s}
\label{massradius}
\end{equation}
introduced in \S \ref{angmomtransport}, and describe the mass
increase due to accretion by  
\begin{equation}
\MWD = M_0\left(1 + {t \over t_{ac}}\right),
\label{Mincrease}
\end{equation}
$M_0$ being the initial mass of the WD, and $R_0$ its 
initial radius.

Combining Equations (\ref{YLs1}), (\ref{ellKsurf}), 
(\ref{massradius}), and (\ref{Mincrease}), and integrating
with respect to time, we obtain the dimensionless specific
angular momentum at the surface, $\lambda_{surf} \equiv 
\ell_{surf}/\ell_0$ in terms of the fiducial value 
$\ell_0 \equiv \sqrt{gM_0R_0}$, as a function of time in
the form
\begin{equation}
\lambda_{surf}(t) = {2 \over 3-s}\left[(1+x)^{3-s \over 2}
- 1 \right]
\label{ellsurf}
\end{equation} 
Here, $x \equiv t/t_{ac}$ is a dimensionless time, in units
of the accretion timescale. The surface angular velocity,
$\Omega_{surf} = \ell_{surf}/\RWD^2$ is similarly given in
terms of the fiducial value $\Omega_0 \equiv \sqrt{GM_0/
R_0^3}$ as
\begin{equation}
{\Omega_{surf}(t) \over \Omega_0}= {2 \over 3-s}
(1+x)^{2s}\left[(1+x)^{3-s \over 2} - 1 \right].
\label{Omsurf}
\end{equation} 

The above increase with time continues until the the angular 
velocity at the WD surface reaches the {\sl instantaneous} 
Keplerian value $\Omega_K{\RWD} \equiv \sqrt{G\MWD/\RWD^3}$ there. 
This happens when the ratio
\begin{equation}
\omega_{surf}(t) \equiv {\Omega_{surf}(t) \over \Omega_K(\RWD)}
= {2 \over 3-s}(1+x)^{s-1 \over 2}
\left[(1+x)^{3-s \over 2} - 1 \right].
\label{omsurf}
\end{equation} 
reaches unity, which occurs at a critical value of the 
(dimensionless) time $x_c$ given by the solution of the equation
\begin{equation}
x - (1+x)^{s-1 \over 2} = {1-s \over 2}.
\label{xcrit}
\end{equation}
For values of $s$ in the range $\sim 1-2$, as is the case for
some actual WD models in the literature (see \S \ref{transport1}), 
the value of $x_c$ is close to 1, as can be easily verified.  

Beyond this point, the surface angular velocity is maintained 
at the current Keplerian angular velocity, \ie, $\omega_{surf}$ 
is maintained at a value of unity. Since the ratio of 
$\lambda_{surf}(t)$ to $\omega_{surf}(t)$ is seen to be
\begin{equation}
{\lambda_{surf}(t) \over \omega_{surf}(t)} 
= (1+x)^{1-s \over 2}
\label{ratio}
\end{equation} 
by combining Equations (\ref{ellsurf}) and (\ref{omsurf}), it 
follows that, under these circumstances, $\ell_{surf}$ is  
given as a function of time by
\begin{equation}
\lambda_{surf}(t) = (1+x)^{1-s \over 2}.
\label{ellsurf1}
\end{equation}

%% \bibliographystyle{apj}
%% \bibliography{bibfile}

\end{document}